\documentclass[prb]{revtex4}
\usepackage{amssymb}
\usepackage{graphicx,latexsym,natbib}

\begin{document}

\title{ Bosonic Resonating Valence Bond wave function for doped Mott
insulators}
\author{Z. Y. Weng}
\affiliation{Center for Advanced Study, Tsinghua University, Beijing 100084, China}
\author{Y. Zhou}
\affiliation{Max-Planck-Institut fur Physik komplexer Systeme \\
Noethnitzer Str. 38, 01187 Dresden, Germany}
\author{V. N. Muthukumar}
\affiliation{Department of Physics, \\
City College at the City University of New York,\\
New York, NY 10031}
\date{today}

\begin{abstract}
We propose a class of ground states for doped Mott insulators in the
electron second quantization representation. They are obtained from a
bosonic resonating valence bond (RVB) theory of the $t-J$ model. At half
filling, the ground state describes spin correlations of the S=1/2
Heisenberg model very accurately. Its spin degrees of freedom are
characterized by RVB pairing of spins, the size of which decreases
continuously as holes are doped into the system. Charge degrees of freedom
emerge upon doping and are described by twisted holes in the RVB background.
We show that the twisted holes exhibit an off diagonal long range order
(ODLRO) in the pseudogap ground state, which has a finite pairing amplitude,
but is short of phase coherence. Unpaired spins in such a pseudogap ground
state behave as free vortices, preventing superconducting phase coherence.
The existence of nodal quasiparticles is also ensured by such a hidden ODLRO
in the ground state, which is non Fermi liquid-like in the absence of
superconducting phase coherence. Two distinct types of spin excitations can
also be constructed. The superconducting instability of the pseudogap ground
state is discussed and a d-wave superconducting ground state is obtained.
This class of pseudogap and superconducting ground states unifies
antiferromagnetism, pseudogap, superconductivity, and Mott physics into a
new state of matter.
\end{abstract}

\maketitle

\section{Introduction}

An important question posed by the study of high temperature superconductors
is whether the superconducting ground state of doped Mott insulators can be
described by a BCS-like wave function. A $d$-wave BCS wave function and/or
its derivatives have been widely used in the phenomenology of high
temperature superconductors, partly because electron pairing, as evidenced
by $2e$ flux quantization and similar experimental results, is observed in
the cuprate superconductors. However, owing to the effect of strong on-site
Coulomb repulsion between the electrons, superconductivity in these systems
may be quite different. Whereas in a conventional BCS superconductor,
screening and retardation effects serve to minimize the role of Coulomb
interaction, the role of the latter becomes crucial in a doped Mott
insulator where the charge degrees of freedom are partially frozen by the
severity of the Coulomb interaction. Strong arguments have been presented in
the literature that a high temperature superconductor evolves from a Mott
insulator doped by holes, and consequently the ground state can be very
different from a conventional BCS state \cite{pwa_87}.

A simple step towards incorporating Mott physics into a BCS description was
suggested many years ago \cite{pwa_87,bza_87}. This involves a projection
(out of the Hilbert space) of doubly occupied sites in a $d$-wave BCS state
\cite{gros_88}. In this approach, the charge degrees of freedom on a lattice
site $i$, are partially frozen by a Gutzwiller operator $(1-n_{i\uparrow
}n_{i\downarrow })$. At half filling, only spin degrees of freedom survive,
and the wave function describes a Mott insulating state. This class of wave
functions has achieved some success in the phenomenology of the cuprate
superconductors \cite{zhang_88, yokoyama_96,paremakanti_01}. A recent paper
argues the case further \cite{pwa_03}. However, the treatment of low energy
spin degrees of freedom in this description is not satisfactory.
Experimentally, strong antiferromagnetic (AF) correlations have been
observed for the Cu spins, but projected $d$-wave BCS states cannot account
correctly for such correlations. Since we opine that spin correlations are
intimately related to superconductivity in a doped Mott insulator, we are
motivated to seek an alternate description in which the spin degrees of
freedom can be treated more accurately and systematically.

In this paper, we propose a new class of ground states based on a
microscopic treatment of the $t-J$ model. These ground state wave functions
are different from the projected BCS wave functions, and possess the
following properties: (i) the spin degrees of freedom can fully restore the
AF long range order (AFLRO) as well as low-lying spin wave excitations at
half filling, and evolve into a spin liquid with strong short-range AF
correlations at low doping; (ii) the no double occupancy constraint is
always satisfied for any doping, and is incorporated self-consistently
instead of being enforced by a \textquotedblleft
brute-force\textquotedblright\ projection; (iii) there is an inherent $d$%
-wave superconducting instability at finite doping. Thus, these wave
functions offer a unified description of antiferromagnetism and
superconductivity in a doped Mott insulator.

To motivate the form of the ground state wave function, let us first
consider the ground state of the undoped Mott insulator on a square lattice.
It is well known that the dynamics of the low energy degrees of freedom in
this case are described by the $S=1/2$ Heisenberg Hamiltonian. While the
exact ground state of this model is not known, a comparison with numerical
results shows that the best variational state, $|\Psi _{0}\rangle $, is
given by \cite{lda_88}
\begin{equation}
|\Psi _{0}\rangle =\sum_{i\in Aj\in B}W(i_{1}-j_{1})\ldots
W(i_{n}-j_{n})(i_{1}j_{1})\ldots (i_{n}j_{n})~.  \label{lda}
\end{equation}%
The ground state (\ref{lda}) is defined to be the bosonic RVB state of the
insulator \cite{lda_88}. Here, $(ij)$ stands for a singlet spin pairing
between opposite sublattice sites $i$ and $j$, and $W(ij)$, the positive
weight factor associated with it. The motivation for such a construction is
to ensure that the ground state satisfies the so-called Marshall sign rule
\cite{marshall_55}. Marshall showed that the ground state of the $S=1/2$
Heisenberg Hamiltonian on a square lattice satisfies the condition, $\mathrm{%
sgn}\left[ \Psi _{0}(\sigma _{1},\sigma _{2},...,\sigma _{N})\right]
=(-1)^{P(\sigma _{1},\sigma _{2},...,\sigma _{N})}$. Here, $P(\sigma
_{1},\sigma _{2},...,\sigma _{N})$ denotes, say, the number of down spins on
sublattice $A$ for the spin configuration $|\sigma _{1},\sigma
_{2},...,\sigma _{N}\rangle $. Then, the matrix elements of the Heisenberg
Hamiltonian are negative and the ground state energy assumes the form of a
sum over negative terms \cite{sutherland_88}. Since each singlet bond $(ij)$
satisfies the Marshall sign, it follows that the wave function of (\ref{lda}%
), which is given by $\Psi _{0}\propto \sum \prod\nolimits_{\left( ij\right)
}(-1)^{i}W(i-j)$ [here $i$($j$) stands for up (down) spin sites and $%
h(i-j)=0 $ if $ij\in $ the same sublattice sites], always satisfies this
criterion.

The ground state (\ref{lda}) describes both short-range and long-range AF
correlations very accurately as the best variational wave function for the
insulator. It is then reasonable to expect that any putative RVB state, away
from half filling, evolves from (\ref{lda}). With this in mind, we ask how
the wave function (\ref{lda}) is modified in the presence of holes. As
discussed above, singlet pairing is characterized in the wave function (\ref%
{lda}) by the Marshall sign. It can be shown that the Marshall sign would be
still obeyed if doped holes remain static. Then, from the form of (\ref{lda}%
), one may easily see that the nearest-neighbor (nn) hopping of holes, which
displaces spins as backflow, generally leads to disordering of the Marshall
sign $(-1)^P$. Consequently, a hole moving along a closed path picks up a
generalized Berry's phase, which was first identified by one of us and
co-authors\cite{weng_97}. This effect, described (and called) as a phase
string is illustrated in Fig. 1(a). It is a singular effect and cannot be
repaired by low energy (transverse) spin excitations \cite{weng_97}.

\begin{figure}[tbph]
\centering\includegraphics{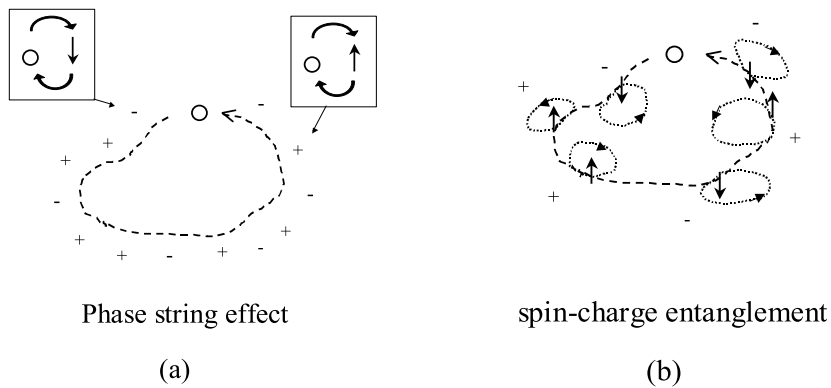}
\caption{(a) The ground state picks up a sequence of signs, $(-1)\times
(+1)\times (+1)\times ...$, known as a phase string, during an operation as
a hole moves through a closed path and back to the origin with the displaced
spin configurations being restored. This irreparable phase string effect has
been shown to be a general consequence of the $t-J$ model \protect\cite%
{weng_97}. (b) The $+$ or $-$ sign in a phase string is determined by the
exchange of the hole with an $\uparrow $ or $\downarrow $ spin on the path,
which can be viewed as an entanglement between the hole and a spin,
illustrated by two entangled loops as both the hole and a displaced spin
should return to their origins after the operation.}
\end{figure}

It follows that a \emph{bare} hole created in the RVB background (\ref{lda})
cannot possibly describe a low-energy state, and must be dressed by the
nonlocal phase string effect. Since the phase string effect describes a
mutual entanglement between spin and charge degrees of freedom illustrated
in Fig. 1(b), the RVB background must also respond nonlocally. One may thus
surmise that low-energy charge degrees of freedom are created by $\tilde{c}%
_{i\sigma }=c_{i\sigma }e^{-i\hat{\Theta}_{\sigma }}$, instead of by the
bare $c_{i\sigma }$, where $e^{-i\hat{\Theta}_{\sigma }}$ is a nonlocal
operator describing the phase string effect. We call a hole that is created
by $\tilde{c}_{i\sigma }$ a twisted hole, and as will be shown in this
paper, such a twisted hole is a bosonic object whose kinetic energy is
optimized in the (modified) RVB background. We shall obtain a ground state
of the following form%
\begin{equation}
|\Psi _{G}\rangle =\left( \sum_{l}Z_{\uparrow }(l)\tilde{c}_{l\uparrow
}\right) ^{N_{h}^{\uparrow }}\left( \sum_{l}Z_{\downarrow }(l)\tilde{c}%
_{l\downarrow }\right) ^{N_{h}^{\downarrow }}|\mathrm{RVB}\rangle ~,
\label{gswf}
\end{equation}%
where $Z_{\sigma }(l)$ describes the wave function of the twisted hole $%
\tilde{c}_{l\sigma }$, and $N_{h}^{\uparrow }+N_{h}^{\downarrow }=N_{h}$ is
the total number of holes. The modified RVB vacuum $|\mathrm{RVB}\rangle $
has the same structure as $|\Psi _{0}\rangle $ at half-filling, except that
the RVB pairing amplitudes, $W_{ij}$'s, become complex and change
continuously with doping, which can be determined self-consistently as we
shall show.

The ground state (\ref{gswf}) is one of the central results of this paper.
It generalizes the wave function (1) written down by Liang \textit{et al.},
to the doped case. We will derive (\ref{gswf}) from a microscopic treatment
of the $t-J$ model, based on a reformulation of the model in the so-called
phase string representation.

The wave function (\ref{gswf}) differs from the Gutzwiller projected BCS
wave functions in a fundamental way. Spin correlations with various length
scales emerge self-consistently as a function of doping in $|\mathrm{RVB}%
\rangle $ and their interplay with the charge degrees of freedom is
incorporated through the nonlocal phase string operator $e^{-i\hat{\Theta}%
_{\sigma }}$ in $\tilde{c}_{i\sigma }$. We will find that the ground state (%
\ref{gswf}) describes both an AF insulator (at zero doping) and a pseudogap
phase with short range AF correlations (at finite doping) in which two
distinct kinds of spin excitations emerge. We will identify a hidden ODLRO
in this pseudogap ground state which ensures a finite pairing amplitude and
nodal fermionic quasiparticles. Although superconducting phase coherence of
this state is prevented by the presence of free spins which behave like
vortices, there exists an inherent superconducting instability at finite
doping, which can naturally give rise to $d$-wave superconductivity.

The outline of the paper follows. In Section II, we reformulate the $t-J$
model using the phase string representation. This is an exact reformulation
and allows us to sort out the effects of hole motion in an AF background
discussed earlier. A singular gauge transformation automatically ensures
that the single occupancy constraint is satisfied. A brief review of this
formalism and important physical consequences will be discussed in Section
II. In Section III, we obtain the ground state within a mean field theory,
and in Section IV we transform the ground state wave function back to the
electron representation, thus completing the derivation of the wave function
(\ref{gswf}). A discussion of the properties predicted by (\ref{gswf})
follows and finally a superconducting ground state is obtained. The final
section is devoted to a general discussion about the region of validity (in
the phase diagram) of this approach, as well as the questions that this
study raises. In Appendix A, some detailed derivations of the mean-field
theory in the phase string representation are presented. In Appendix B, we
discuss briefly, an effective theory of spin and charge degrees of freedom
that are coupled by the phase string. In Appendix C, we show that the ground
state obeys time reversal and spin rotational symmetries.

\section{Phase string formulation and physical implications}

In this section, we reformulate the $t-J$ model using the phase string
representation \cite{weng_97}. This representation, like the slave-particle
representations used in the literature \cite{zou_88, sfermion}, is an exact
reformulation. However, the phase string formulation has a distinct
advantage over these other schemes, in the following sense. The most
singular effects encapsulated in the $t-J$ model are sorted out by a unitary
transformation, and all nontrivial phases present in the original model now
occur as topological gauge fields. Without these gauge fields, the model
becomes trivial, \emph{i.e.}, free of sign problems, which is the case in
one dimension, or in the particular case of half filling in two dimensions.
The central issue in two dimensions is then how to handle these topological
gauge fields at finite doping. It turns out that they are well controlled,
at least for small doping.

\subsection{ Phase string representation}

The most singular effect of the hole moving in an AF background is the phase
string effect. This leads to a competition between hopping and superexchange
processes. Since it is most singular at length scales of a lattice constant
(as Z$_2$ signs), it is crucial to track this effect explicitly, before
constructing any effective theory. As shown by Weng \textit{et al.}, this
can be achieved starting from a slave-fermion representation and by
performing a unitary transformation of the basis states \cite{weng_97}. In
this procedure, the electron operator is decomposed into
\begin{equation}
c_{i\sigma }=h_i^{\dagger }b_{i\sigma }e^{i\hat{\Theta}_{i\sigma }}
\label{mutual}
\end{equation}
where holon $h_{i\sigma }^{\dagger }$ and spinon $b_{i\sigma }$ operators
are both bosonic fields, satisfying the constraint
\begin{equation}
h_i^{\dagger }h_i+\sum_\sigma b_{i\sigma }^{\dagger }b_{i\sigma }=1~.
\label{constr}
\end{equation}
The phase factor $e^{i\hat{\Theta}_{i\sigma }}$ in Eq. (\ref{mutual}) is
defined by
\begin{equation}
e^{i\hat{\Theta}_{i\sigma }}\equiv e^{i\Theta _{i\sigma }^{string}}~(\sigma
)^{\hat{N}_h}(-\sigma )^i,  \label{phase}
\end{equation}
where $\hat{N}_h$ is the (total) holon number operator, the phase string
operator $\Theta _{i\sigma }^{string}$ $\equiv \frac 12\left[ \Phi _i^s-\Phi
_i^0-\sigma \Phi _i^h\right] $ is a nonlocal operator with
\begin{eqnarray}
\Phi _i^s &=&\sum_{l\neq i}\theta _i(l)\left( \sum_\alpha \alpha n_{l\alpha
}^b\right) ~,  \label{phis} \\
\Phi _i^0 &=&\sum_{l\neq i}\theta _i(l)~,  \label{phi0}
\end{eqnarray}
and
\begin{equation}
\Phi _i^h=\sum_{l\neq i}\theta _i(l)n_l^h~.  \label{phih}
\end{equation}
Here, $n_{l\alpha }^b$ and $n_l^h$ are spinon and holon number operators
respectively, at site $l,$ and $\theta _i(l)\equiv \mbox{Im ln $(z_i-z_l)$},$
where $z_l=x_l+iy_l$ is a complex coordinate on the lattice.

It is easily verified that the fermionic statistics of $c_{i\sigma }$ is
automatically ensured by $e^{i\Theta _{i\sigma }^{string}}$; besides
ensuring fermionic statistics, the string operator also incorporates the
singular effects of hole motion in the RVB background, as will be discussed
later. The factor $(\sigma )^{\hat{N}_{h}}$ guarantees anticommutation
relations between opposite spins. The factor $(-\sigma )^{i}=\pm 1$ in (\ref%
{phase}) is added for convenience, and incorporates the Marshall sign into
the decomposition (\ref{mutual}) of the electron in terms of the holon and
spinon operators, which may be regarded as a bosonization scheme for the
electron moving in an RVB background. The same bosonization decomposition
can be also obtained from the slave-boson representation \cite%
{weng_94,qhwang}.

\subsection{The $t-J$ model in the phase string representation}

Rewriting the $t-J$ model using the electron decomposition (\ref{mutual}),
we get $H_{t-J}=H_t+H_J,$ where the hopping term
\begin{equation}
H_t=-t\sum_{\langle ij\rangle \sigma }\left( e^{iA_{ij}^s-i\phi
_{ij}^0}\right) h_i^{\dagger }h_j\left( e^{i\sigma A_{ji}^h}\right)
b_{j\sigma }^{\dagger }b_{i\sigma }+h.c.~ ,  \label{ht}
\end{equation}
and the superexchange term 
\begin{equation}
H_J=-\frac J2\sum_{\langle ij\rangle }~\left( \hat{\Delta}_{ij}^s\right)
^{\dagger }\hat{\Delta}_{ij}^s~.  \label{hj}
\end{equation}
The RVB pair operator $\hat{\Delta}_{ij}^s$ in $H_J$ is defined by
\begin{equation}
\hat{\Delta}_{ij}^s\equiv \sum_\sigma e^{-i\sigma A_{ij}^h}b_{i\sigma
}b_{j-\sigma }~.  \label{brvb}
\end{equation}
The unique feature in this formalism is the emergence of three phases (link
fields): $A_{ij}^s$, $\phi _{ij}^0$, and $A_{ij}^h,$ defined on the nn
links. Without these, there should be \emph{no} nontrivial sign problem in $%
H_{t-J}$, because $h$ and $b$ are both bosonic fields. The matrix elements
of $H_{t-J}$ would then be real and negative-definite in the occupation
number representation of $h$ and $b$. Consequently, the ground state
expanded in terms of these bosonic fields will have real and positive
coefficients. Therefore, any nontrivial signs (phases) of the $t-J$ model
arise solely from these three link fields, defined by
\begin{equation}
A_{ij}^s\equiv \frac 12\sum_{l\neq i,j}\left[ \theta _i(l)-\theta _j(l)%
\right] \left( \sum_\sigma \sigma n_{l\sigma }^b\right) ~,  \label{as}
\end{equation}
\begin{equation}
\phi _{ij}^0\equiv \frac 12\sum_{l\neq i,j}\left[ \theta _i(l)-\theta _j(l)%
\right] ~,  \label{sphi0}
\end{equation}
and
\begin{equation}
A_{ij}^h\equiv \frac 12\sum_{l\neq i,j}\left[ \theta _i(l)-\theta _j(l)%
\right] n_l^h~.  \label{ah}
\end{equation}

It is easy to see that the Hamiltonian $H_{t-J}$ is invariant under \textrm{%
U(1)}$\times $\textrm{U(1)} gauge transformations:
\begin{equation}
h_i\rightarrow h_ie^{i\varphi _i}~,\text{ \qquad \quad }A_{ij}^s\rightarrow
A_{ij}^s+(\varphi _i-\varphi _j)~,  \label{u(1)1}
\end{equation}
and
\begin{equation}
b_{i\sigma }\rightarrow b_{i\sigma }e^{i\sigma \theta _i}~,\text{ \qquad }%
A_{ij}^h\rightarrow A_{ij}^h+(\theta _i-\theta _j)~.  \label{u(1)2}
\end{equation}
Thus $A_{ij}^s$ and $A_{ij}^h$ are gauge fields, seen by holons and spinons
respectively, as the latter carry their gauge charges according to (\ref%
{u(1)1}) and (\ref{u(1)2}). But they are not conventional \textrm{U(1)}
gauge fields, as they must satisfy the topological constraints: $%
\sum_cA_{ij}^s=\pi \sum_{l\in c}\left( n_{l\uparrow }^b-n_{l\downarrow
}^b\right) $ and $\sum\nolimits_cA_{ij}^h=\pi \sum_{l\in c}n_l^h$ according
to (\ref{as}) and (\ref{ah}), respectively. Here the subscript $c$ denotes
an arbitrary closed loop such that the fluxes enclosed, $\sum_c$ $A_{ij}^s$
and $\sum\nolimits_cA_{ij}^h$, are determined by the number of spinons and
holons respectively, inside the loop $c$; \emph{viz.}, $A_{ij}^s$ and $%
A_{ij}^h$ are not independent gauge fields with their own dynamics. Rather
they are directly connected to matter fields as a pair of mutual
Chern-Simons gauge fields \cite{kqw}. The term ``mutual'' refers to the fact
that $A_{ij}^s$ describes quantized $\pi $ fluxoids attached to spinons,
coupled to the holons in $H_t$. Conversely, $A_{ij}^h$ describes quantized $%
\pi $ fluxoids bound to holons, coupled to spinons in $H_J$. The phase $\phi
_{ij}^0$ describes a uniform and constant flux with a strength $\pi $ per
plaquette: $\sum_{\Box }\phi _{ij}^0=\pm \pi $ according to (\ref{sphi0}).

In the case of the one dimensional model, $A_{ij}^s=\phi _{ij}^0=A_{ij}^h=0.$
Thus there is no sign problem in the phase-string representation of the $t-J$
model. It implies that the Hamiltonian can be treated within a mean field
framework, which indeed results in a correct Luttinger-liquid solution for
the large-U Hubbard model \cite{weng_97}. In two dimensions, these nn link
phases can no longer be ``gauged away''. They compose a mutual gauge
structure which completely determines the essential sign problem of the $t-J$
model. These gauge fields are generally well controlled in the regimes of
our interest: $\phi _{ij}^0$ is a non-dynamic phase describing a constant $%
\pi $ flux per plaquette; $A_{ij}^s$ is canceled when spinons are RVB paired
at low-temperature phases; $A_{ij}^h$ remains weak at small doping - it
vanishes at half filling, where there is no sign problem in the Hamiltonian.
Therefore, these gauge fields in the phase string representation are
presumably well tractable in low doping and low temperature regimes. This is
one of the main advantages of the phase string representation over other
approaches.

Finally, it is important to note that the phase string representation is
defined in a Hilbert space where the total $S^z$ is an eigen operator. The
total numbers of $\uparrow $ and $\downarrow $ spinons are conserved
individually, such that the topological gauge field $A_{ij}^s$ behaves
smoothly as defined in (\ref{as}), depicting fictitious $\pi $ fluxoids
bound to spinons. (Non-conserved $S^z$ would result in non-conserved $+$ and
$-$ fluxoids in $A_{ij}^s$). Different $S^z$ states are connected by the
spin flip operators, defined in the phase-string representation as
\begin{equation}
S_i^{+}=\left[ (-1)^ie^{i\Phi _i^h}\right] b_{i\uparrow }^{\dagger
}b_{i\downarrow },  \label{s+}
\end{equation}
(a factor $(-1)^{\hat{N}_h}$ has been dropped for simplicity) and $%
S_i^{-}=(S_i^{+})^{\dagger }$, and $S_i^z=\sum_\sigma \sigma b_{i\sigma
}^{\dagger }b_{i\sigma }$. These definitions follow from (\ref{mutual}). The
nonlocal phase $\Phi _i^h$ in (\ref{s+}) plays a crucial role in restoring
the spin rotational symmetry.

\subsection{No double occupancy constraint}

Since the $t-J$ model is a projective Hamiltonian, the no double occupancy
constraint is a central issue. It imposes a severe restriction on the
Hilbert space
\begin{equation}
\sum_\sigma c_{i\sigma }^{\dagger }c_{i\sigma }\leq 1,
\end{equation}
which ensures that the system is a Mott insulator at half filling and a
doped Mott insulator away from half filling. In conventional slave particle
decompositions, the constraint reduces to a single occupancy constraint like
(\ref{constr}), and is still difficult to treat. Typically, the constraint
is relaxed at the level of mean field theory, and restoring its effect
beyond the mean field theory (say, by a gauge theory) is a very challenging
problem. In contrast, the phase string representation outlined in the
previous subsection provides a new way to handle the constraint.

A wavefunction $\psi _e$ in the electron $c$-operator representation is
defined in a quantum state as
\begin{equation}
|\Psi \rangle =\sum_{\{i_u\}\{j_d\}}\psi _e(\{i_u\};\{j_d\})\text{ }%
c_{i_1\uparrow }^{\dagger }c_{i_2\uparrow }^{\dagger }\cdot \cdot \cdot
c_{i_M\uparrow }^{\dagger }c_{j_1\downarrow }^{\dagger }c_{j_2\downarrow
}^{\dagger }\cdot \cdot \cdot c_{j_{N_e-M}\downarrow }^{\dagger }|0\rangle ~,
\label{psie}
\end{equation}
where the $\uparrow $ spin electron sites, $\{i_u\}=i_1,\cdot \cdot \cdot
,i_M,$ and the $\downarrow $ spin sites, $\{j_d\}=j_1,\cdot \cdot \cdot
,j_{N_e-M},$ obey the no double occupancy constraint in the restricted
Hilbert space of the $t-J$ Hamiltonian. On the other hand, in the phase
string representation, the wave function $\psi _b(i_1,\cdot \cdot \cdot
,i_M;j_1,\cdot \cdot \cdot ,j_{N_e-M};l_1,\cdot \cdot \cdot ,l_{N_h})$ is
defined as
\begin{equation}
|\Psi \rangle =\sum_{\{i_u\}\{j_d\}}\psi
_b(\{i_u\};\{j_d\};\{l_h\})|\{i_u\};\{j_d\};\{l_h\}\rangle ~,  \label{psib-0}
\end{equation}
where the indices $\{l_h\}=l_1,\cdot \cdot \cdot ,l_{N^h}$ denote the empty
sites, that are \emph{not} independent from $\{i_u\}$ and $\{j_d\}$ under
the constraint. Here the spinon-holon basis in the phase string
representation, $|\{i_u\};\{j_d\};\{l_h\}\rangle $ is given by
\begin{equation}
|\{i_u\};\{j_d\};\{l_h\}\rangle \equiv (-1)^{N_A^{\uparrow }}b_{i_1\uparrow
}^{\dagger }b_{i_2\uparrow }^{\dagger }\cdot \cdot \cdot b_{i_M\uparrow
}^{\dagger }b_{j_1\downarrow }^{\dagger }b_{j_2\downarrow }^{\dagger }\cdot
\cdot \cdot b_{j_{N^e-M}\downarrow }^{\dagger }h_{i_1}h_{i_2}\cdot \cdot
\cdot h_{j_{N_e-M}}|0\rangle ~,  \label{b-basis}
\end{equation}
where the vacuum $|0\rangle $ is chosen to be filled by holons. The sign
factor $(-1)^{N_A^{\uparrow }}$ in (\ref{b-basis}) can be identified with
the Marshall sign, and $N_A^{\uparrow }$ denotes the total number of $%
\uparrow $ spins in sublattice $A$. Here and henceforth, we will always use $%
i$ to specify an $\uparrow $ spin, $j$ a $\downarrow $ spin, and $l$, a
holon, where the subscripts $u$, $d$, and $h$ label the sequences of the $%
\uparrow $ spins, $\downarrow $ spins, and holons, respectively.

Now, the decomposition (\ref{mutual}) of the electron operator $c_{i\sigma }$%
, relates the wave functions written in the two representations:
\begin{equation}
c_{i_1\uparrow }^{\dagger }c_{i_2\uparrow }^{\dagger }\cdot \cdot \cdot
c_{i_M\uparrow }^{\dagger }c_{j_1\downarrow }^{\dagger }c_{j_2\downarrow
}^{\dagger }\cdot \cdot \cdot c_{j_{N^e-M}\downarrow }^{\dagger }|0\rangle =%
\mathcal{K}^{-1}|\{i_u\};\{j_d\};\{l_h\}\rangle ~,  \label{relation}
\end{equation}
where
\begin{eqnarray}
\mathcal{K} &=&\prod_{ud}\frac{z_{i_u}^{*}-z_{j_d}^{*}}{|z_{i_u}-z_{j_d}|}%
\prod_{u<u^{\prime }}\frac{z_{i_u}^{*}-z_{i_{u^{\prime }}}^{*}}{%
|z_{i_u}-z_{i_{u^{\prime }}}|}\prod_{d<d^{\prime }}\frac{%
z_{j_d}^{*}-z_{j_{d^{\prime }}}^{*}}{|z_{j_d}-z_{j_{d^{\prime }}}|}\prod_{uh}%
\frac{z_{i_u}^{*}-z_{l_h}^{*}}{|z_{i_u}-z_{l_h}|}  \nonumber \\
&=&\mathcal{C}^{-1}\prod_{ud}(z_{i_u}^{*}-z_{j_d}^{*})\prod_{u<u^{\prime
}}(z_{i_u}^{*}-z_{i_{u^{\prime }}}^{*})\prod_{d<d^{\prime
}}(z_{j_d}^{*}-z_{j_{d^{\prime }}}^{*})\prod_{h<h^{\prime
}}|z_{l_h}-z_{l_{h^{\prime }}}|\left(
\prod_{uh}(z_{i_u}^{*}-z_{l_h}^{*})\prod_{dh}|z_{j_d}-z_{l_h}|\right) ~.
\label{k0}
\end{eqnarray}
The coefficient $\mathcal{C}$ is given by
\begin{eqnarray*}
\mathcal{C} &=&\prod_{ud}|z_{i_u}-z_{j_d}|\prod_{u<u^{\prime
}}|z_{i_u}-z_{i_{u^{\prime }}}|\prod_{d<d^{\prime }}|z_{j_d}-z_{j_{d^{\prime
}}}|\prod_{h<h^{\prime }}|z_{l_h}-z_{l_{h^{\prime
}}}|\prod_{uh}|z_{i_u}-z_{l_h}|\prod_{dh}|z_{j_d}-z_{l_h}| \\
&\equiv &\prod_{k<m}|z_k-z_m|~,
\end{eqnarray*}
in which $k$ and $m$ run through all lattice sites, such that $\mathcal{C}$
is a constant.

Correspondingly, the wave functions in the electron and phase string
representations are related by
\begin{equation}
\psi _e(i_1,\cdot \cdot \cdot ,i_M;j_1,\cdot \cdot \cdot ,j_{N_e-M})=
\mathcal{K} \text{ }\psi_b(i_1,\cdot \cdot \cdot ,i_M;j_1,\cdot \cdot \cdot
,j_{N_e-M};l_1,\cdot \cdot \cdot ,l_{N_h})~,  \label{wavefunction}
\end{equation}
which holds generally for the $t-J$ model.

Note that the above expression for $\mathcal{K}$ has been obtained strictly
under the no double occupancy constraint. But if the constraint is relaxed
in (\ref{k0}), while still treating $\mathcal{C}$ as a constant, then one
finds $\mathcal{K}=0$ for double occupancies when \emph{any }two particles
(spinons and/or holons) occupy the same site; \emph{viz.}, $\mathcal{K}$
defined in (\ref{k0}) automatically enforces the constraint (\ref{constr})
through the Jastrow-like factors. Therefore, as far as $\psi _e$ is
concerned, the no double occupancy constraint in $\psi _b$ or $\psi _b/
\mathcal{C}$ is no longer important, since $\mathcal{K}$ in (\ref%
{wavefunction}) naturally plays the role of a projection operator.

Clearly, an exact $\psi _b$ would satisfy the single occupancy constraint (%
\ref{constr}). But, the point is that an approximate $\psi _b$ determined
without the constraint does not affect $\psi _e$, owing to $\mathcal{K}$. It
means that in the phase string representation the constraint (\ref{constr})
is indeed ``unimportant'', which may be understood in the following way. In
the phase string representation, the effect of $\mathcal{K}$ in the original
$\psi _e$, is transformed into the topological gauge fields$,$ $A_{ij}^s$
and $A_{ij}^h,$ in the Hamiltonians, (\ref{ht}) and (\ref{hj}), which
describe spinons and holons as mutual vortices, as perceived by each other (%
\textit{cf.} the discussion in the previous subsection). This clearly
implies a mutual repulsion between two species, since a spinon cannot stay
at the center of its vortex (which is a holon), and \textit{vice versa}.
Thus the constraint that a holon and a spinon cannot occupy the same site is
now reflected in the \emph{interactions} present in the new Hamiltonian, and
the condition (\ref{constr}) is not needed as an extra condition to enforce.
Note that the constraint (\ref{constr}) also requires the hard core
conditions among the holons or spinons themselves. But since both holon and
spinon fields are bosonic fields, local hard core exclusions usually do not
involve the sign change of the wave function. Hence, in the phase string
representation, the local constraint (\ref{constr}) is neither crucial nor
singular, as far as low energy physics is concerned.

\subsection{Phase string effect: A singular doping effect}

In this subsection, we point out that the other important effect, \emph{i.e.}
, the phase string effect, is also explicitly built into this representation
via $\mathcal{K}$.

Let us first rewrite $\mathcal{K}$ (\ref{k0}) in a more compact form,
\begin{equation}
\mathcal{K}=\mathcal{JG}~,  \label{k2}
\end{equation}%
where
\begin{equation}
\mathcal{J}\equiv \prod_{u<u^{^{\prime }}}(z_{i_{u}}^{\ast }-z_{i_{u^{\prime
}}}^{\ast })\prod_{d<d^{^{\prime }}}(z_{j_{d}}^{\ast }-z_{j_{d^{\prime
}}}^{\ast })\prod_{ud}(z_{i_{u}}^{\ast }-z_{j_{d}}^{\ast
})\prod_{h<h^{\prime }}\left\vert z_{l_{h}}-z_{l_{h^{\prime }}}\right\vert
\prod_{uh}|z_{i_{u}}-z_{l_{h}}|\prod_{dh}|z_{j_{d}}-z_{l_{h}}|~,  \label{jh}
\end{equation}%
and
\begin{equation}
\mathcal{G}\equiv \mathcal{C}^{-1}\prod_{uh}\frac{z_{i_{u}}^{\ast
}-z_{l_{h}}^{\ast }}{|z_{i_{u}}-z_{l_{h}}|}~.  \label{g0}
\end{equation}%
It is easily seen that the Jastrow-like factors in $\mathcal{J}$ enforce the
single occupancy constraint discussed previously: $\mathcal{J}$ vanishes if
two spins (or holes) occupy the same site, or if a hole and a spin occupy
the same site. The factor $\mathcal{J}$ also explicitly captures the
fermionic statistics of electrons.

Now let us focus on the additional factor $\mathcal{G}$ in $\mathcal{K}$,
which is asymmetric with regard to $\uparrow $ and $\downarrow $ spins: it
only involves an $\uparrow $ spin complex coordinate $z_{i_u}^{*}$ and a
holon coordinate $z_{l_h}^{*}$. Suppose that the holon is taken along a
closed loop shown in Fig. 1. The $\mathcal{K}$ factor will generally acquire
a nontrivial phase through $\mathcal{G}$. In the following we examine this.

For a spinon inside or outside the loop, there will be no net contribution
to $\mathcal{G}$ as the acquired phase is either $2\pi $ or $0$ under such
an operation. So any nontrivial contribution is solely from spinons
occupying lattice sites on the holon's loop-path. At each step of hopping,
the holon under consideration will have to switch positions with a spinon
occupying a nn site on the loop, obeying the no double occupancy constraint.
Then $\mathcal{G}$ picks up a minus sign from the Jastrow factor in $%
\mathcal{G}$ every time when the holon and an $\uparrow $ spinon are
exchanged. Thus, if one changes the holon's coordinate continuously through
the loop without encountering other holons, $\mathcal{G}$ will acquire a
sequence of signs,
\begin{equation}
\mathcal{G\longrightarrow }\text{ }\left( -1\right) ^{N_{\text{loop}%
}^{\uparrow }}\mathcal{G},  \label{pstring}
\end{equation}%
where $N_{\text{loop}}^{\uparrow }$ denotes the number of $\uparrow $
spinons exchanged with the holon during the above thought experiment. (Note
that the spinons displaced by holon hopping during the operation should all
be restored to the original configuration by \emph{pure} spin flips in terms
of the superexchange $H_{J}$, and this does not generate any further signs
in $\mathcal{G})$.

Therefore, the electron wave function $\psi _e$, acquires a sign factor
(which can also be thought of as a Berry phase) through $\mathcal{G}$:
\begin{equation}
\left( -1\right) ^{N_{\text{loop}}^{\uparrow }}=\prod_{\mathrm{loop}
}(-\sigma _m)=\prod_{\mathrm{loop}}\sigma _m~.  \label{pstring1}
\end{equation}
This factor (for the closed loop) is just the phase string shown in Fig. 1,
where $\sigma _m$ denotes the index of spins on the loop exchanged with a
holon, under the thought experiment of moving the holon around a closed path
(note that a closed loop on a bipartite lattice always involves an even
number of nn links).
Equation (\ref{pstring1}) shows that the singular hopping effect is
explicitly incorporated in $\psi _e$ through $\mathcal{K}$, in the phase
string formulation of the $t-J$ model.

To conclude this section, the two singular effects in the $t-J$ model, \emph{%
viz.}, the single occupancy constraint and the phase string effect, are
explicitly represented by the factor $\mathcal{K}$, (\ref{k0}). Consequently
the ground state wave function in the phase string representation, $\psi
_{b},$ is presumably no longer singular, justifying a mean field theory.

\section{Construction of the ground state in phase string representation}

A mean field theory (bosonic RVB theory) based on the phase string
representation of the $t-J$ model was constructed in an earlier paper \cite%
{weng_99}, and we will follow a similar route here in dealing with the spin
degrees of freedom. However, an important distinction in this work is the
way the charge degrees of freedom are incorporated, in order to better
facilitate and optimize the kinetic energy of the holes, as well as to
construct a state in which the no double occupancy constraint is strictly
enforced. As shown in Sec. IV, this construction directly results in a
pseudogap ground state characterized by a finite pairing amplitude and the
emergence of nodal quasiparticles. The ground state has an inherent
superconducting instability which occurs, when phase coherence is realized.

\subsection{Spin degrees of freedom and bosonic RVB order parameter}

Let us begin by considering the superexchange term $H_{J}$, which is
expressed in terms of an RVB pair operator $\hat{\Delta}_{ij}^{s}$ in (\ref%
{hj}). Note that $\hat{\Delta}_{ij}^{s}$ is invariant under the gauge
transformation (\ref{u(1)2}). It is then natural to define the bosonic RVB
order parameter by
\begin{equation}
\Delta ^{s}\equiv \left\langle \hat{\Delta}_{ij}^{s}\right\rangle _{nn}
\label{order}
\end{equation}%
for nn sites, $i$ and $j$. This RVB order parameter is different from the
RVB order parameter introduced by Anderson and collaborators \cite{pwa_87,
bza_87}: it involves the charge degrees of freedom through the gauge field, $%
A_{ij}^{h}$, and is gauge invariant under (\ref{u(1)2}). Note that $%
\left\langle b_{i\sigma }^{\dagger }b_{j\sigma }\right\rangle _{nn}\equiv 0$
in this theory, and $\Delta ^{s}$ is the sole order parameter. At half
filling, $A_{ij}^{h}$ $=0$ in $\hat{\Delta}_{ij}^{s}$, and $\Delta ^{s}$
reduces to the well known Schwinger-boson mean field order parameter $%
\sum_{\sigma }\langle b_{i\sigma }b_{j-\sigma }\rangle $ \cite{sfermion} [in
our definition $b_{i\sigma }$ differs from the conventional Schwinger boson
representation by a sign factor $(-\sigma )^{i}$]. The mean field solution
describes both long and short range AF correlations of the Heisenberg model
quite well, as will be discussed below. Thus, the bosonic RVB theory we
present is ensured to yield good results at half filling, and provides a
natural generalization away from half filling.

We now show how the spin degrees of freedom are treated within the bosonic
RVB theory. The superexchange Hamiltonian is rewritten as
\begin{equation}
H_{s}=-J_{s}\sum_{\langle ij\rangle \sigma }\left( e^{i\sigma
A_{ij}^{h}}\right) b_{i\sigma }^{\dagger }b_{j-\sigma }^{\dagger
}+h.c.~-\lambda \sum_{i}(\sum_{\sigma }b_{i\sigma }^{\dagger }b_{i\sigma
}+\delta -1)~,  \label{hs0}
\end{equation}%
where $J_{s}\equiv J\Delta ^{s}/2$, and $\Delta ^{s}$ is assumed to be real
and positive. The Lagrangian multiplier $\lambda $ is introduced to
implement the single occupancy constraint (\ref{constr}) on the average. We
reemphasize that $H_{s}$ is not a mean field Hamiltonian in the conventional
sense. Spinons are coupled to a gauge field $A_{ij}^{h}$, related to the
charge degrees of freedom, and the Hamiltonian is invariant under the
internal gauge transformation (\ref{u(1)2}). So $H_{s}$ is still a gauge
model, describing low energy, long wavelength spin fluctuations, underpinned
by the bosonic RVB order parameter (\ref{order}). The link field $A_{ij}^{h}$
defined in (\ref{ah}) describes holons as $\pi $ fluxoids perceived by
spinons in (\ref{hs0}). In the ground state, $A_{ij}^{h}$ will be treated as
a constant field, as though the holons are Bose condensed (which condition
will be determined self consistently, when we discuss the charge degrees of
freedom), with
\begin{equation}
\sum_{\square }{A}_{ij}^{h}\simeq \pi \delta  \label{fluxh}
\end{equation}%
for each plaquette, where $\delta $ is the hole concentration.

Then, $H_{s}$ can be diagonalized by a Bogoliubov transformation \cite%
{weng_99}
\begin{equation}
b_{i\sigma }=\sum_{m}w_{m\sigma }(i)\left( u_{m}\gamma _{m\sigma
}-v_{m}\gamma _{m-\sigma }^{\dagger }\right) ~,  \label{bogo}
\end{equation}%
as
\begin{equation}
H_{s}=\sum_{m\sigma }E_{m}\gamma _{m\sigma }^{\dagger }\gamma _{m\sigma }+{%
const}.
\end{equation}%
Here, $u_{m}=\frac{1}{\sqrt{2}}\left( \frac{\lambda }{E_{m}}+1\right)
^{1/2}, $ $v_{m}=\frac{1}{\sqrt{2}}\left( \frac{\lambda }{E_{m}}-1\right)
^{1/2}\mathrm{sgn}(\xi _{m}),$ and $E_{m}=\sqrt{\lambda ^{2}-\xi _{m}^{2}}.$
The Lagrangian multiplier $\lambda $ is determined by enforcing $%
\sum_{i}\sum_{\sigma }\left\langle b_{i\sigma }^{\dagger }b_{i\sigma
}\right\rangle =(1-\delta )N$. The wave function $w_{m\sigma }$ and the
spectrum $\xi _{m}$ are determined by the following eigen equation,
\begin{equation}
\xi _{m}w_{m\sigma }(i)=-J_{s}\sum_{j=nn(i)}e^{i\sigma {A}%
_{ij}^{h}}w_{m\sigma }(j)~.  \label{ew}
\end{equation}

The ground state of $H_s$ is constructed in the usual way by imposing the
condition, $\gamma _{m\sigma }|\mathrm{RVB}\rangle _{\mathrm{MF}}=0$,
\begin{equation}
|\mathrm{RVB}\rangle _{\mathrm{MF}}=\exp \left( \sum_{ij}W_{ij}b_{i\uparrow
}^{\dagger }b_{j\downarrow }^{\dagger }\right) |0\rangle~,  \label{phirvb}
\end{equation}
where the RVB amplitude $W_{ij}$ is given by
\begin{eqnarray}
W_{ij} &\equiv &-\sum_m\frac{v_m}{u_m}w_{m\uparrow }^{*}(i)w_{m\uparrow }(j)
\nonumber \\
&=&-\sum_m\frac{v_m}{u_m}w_{m\downarrow }(i)w_{m\downarrow }^{*}(j)~.
\label{wij}
\end{eqnarray}
In deriving the above, we use $w_{m\sigma }^{*}(i)=w_{m-\sigma }(i)$, which
follows from (\ref{ew}). Furthermore, it is easy to show that for each state
labeled by $m$, there is always a state labeled by $\bar{m}$, for which $\xi
_{\bar{m}}=-\xi _m$ and $w_{\bar{m}\sigma }(i)=(-1)^iw_{m\sigma }(i)$, such
that
\begin{eqnarray}
W_{ij} &=&\frac{1-(-1)^{i-j}}2W_{ij}  \nonumber \\
&\neq &0\text{ \qquad \textrm{only if} }i,j\in \text{\textrm{different
sublattices~, }}  \label{w1}
\end{eqnarray}
\emph{i.e.}, the RVB amplitude $W_{ij}$ only connects $\uparrow $ and $%
\downarrow $ spins on opposite sublattices.

The nn spin correlations are determined by
\begin{equation}
\left\langle \mathbf{S}_{i}\cdot \mathbf{S}_{j}\right\rangle _{nn}=-\frac{3}{%
8}\left| \Delta ^{s}\right| ^{2}<0  \label{s-s}
\end{equation}%
which is antiferromagnetic in nature [\emph{i.e.}, in the resulting bosonic
RVB theory, $\left\langle \mathbf{S}_{i}\cdot \mathbf{S}_{j}\right\rangle >$
0 ($<0$), if $i$ and $j$ belong to the same (opposite) sublattice].
Therefore, $\Delta ^{s}\neq 0$ generally characterizes a regime with short
range AF correlations, present in a wide range of temperature in the $t-J$
model at low doping (e.g., $\sim J/k_{B}\sim 1,500$ K in the zero-doping
limit). Fig. 2 sketches a phase diagram in which $\Delta ^{s}$ or
short-range AF correlations controls the high-energy, short-distance
correlations in a low-doping regime ($<x_{\mathrm{RVB}})$ \cite{weng_99}.
Such a spin singlet-pairing regime, characterized by $\Delta ^{s}\neq 0$,
defines the ``pseudogap''\ regime of the bosonic RVB state, with a
characteristic temperature $T_{0}$, as illustrated in Fig. 2.

In the following two subsections, we further examine the spin correlations
at half-filling and finite doping within this bosonic RVB description.

\begin{figure}[tbph]
\centering \includegraphics{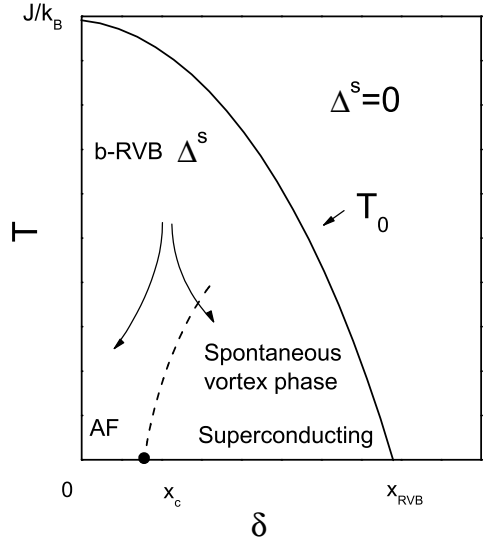}
\caption{Schematic phase diagram of the b-RVB mean field theory: The bosonic
RVB order parameter $\Delta^s$ underpins a pseudogap phase at $T<T_0$, in
which the AF state, spontaneous vortex phase (a low-T pseudogap state), as
well as the superconducting instability occur at low temperatures, separated
by some critical dopings, $x_c$ and $x_{\mathrm{RVB}}$, respectively.}
\end{figure}

\subsubsection{Antiferromagnetism at half-filling}

At half filling, the bosonic RVB ground state $|\mathrm{RVB}\rangle \equiv
P_{G}$ $|\mathrm{RVB}\rangle _{\mathrm{MF}}$, which satisfies the
no-double-occupancy constraint ($P_G$ denotes the Gutzwiller projection
operator), can be explicitly written as
\begin{equation}
|\mathrm{RVB}\rangle =\sum_{\{\sigma _{s}\}}\left( \sum_{\mathrm{pair}%
}\prod_{(ij)}W_{ij}\right) \left( b_{1\sigma _{1}}^{\dagger }h_{1}\right)
\left( b_{2\sigma _{2}}^{\dagger }h_{2}\right) \cdot \cdot \cdot \left(
b_{N\sigma _{N}}^{\dagger }h_{N}\right) |0\rangle ~,  \label{rvb1}
\end{equation}%
where each site is occupied only by one spinon, with \thinspace $\{\sigma
_{s}\}\equiv \sigma _{1},\sigma _{2},\cdot \cdot \cdot ,\sigma _{N}$ being
partitioned into RVB pairs connected by the RVB amplitudes, $W_{ij}$. Here
each RVB pair is denoted by $(ij)$, where $i$($j$) refers to the position of
an $\uparrow $ ($\downarrow $) spin, as before. The summation runs over all
possible RVB pairing partitions. The holon annihilation operators in (\ref%
{rvb1}) are introduced since the spinon vacuum $|0\rangle $ is defined as a
state where every site is occupied by a holon, for later convenience.

$|\mathrm{RVB}\rangle $ given in (\ref{rvb1}) is equivalent to the class of
variational wave functions (\emph{cf} Eq.(\ref{lda})) proposed by Liang,
Doucot, and Anderson (LDA) \cite{lda_88}. The RVB amplitudes can be
determined from the bosonic mean field theory outlined above. At half
filling, $A_{ij}^{h}=0$, and the mean field Hamiltonian (\ref{hs0}) reduces
to the Schwinger-boson mean field Hamiltonian. In this case, the RVB
amplitude $W_{ij}\propto |\mathbf{r}_{ij}|^{-3}$, for $\left\vert \mathbf{r}%
_{ij}\right\vert \gg a$ ($a$ being the lattice constant), as shown in the
inset of Fig.3. We use the loop-gas method employed by LDA to determine the
nn spin correlation (which also decides the ground state energy of the
Heisenberg model), as well as the staggered magnetization, $m$. The maximum
sample size is $64\times 64$. We obtained $\left\langle \mathbf{S}_{i}\cdot
\mathbf{S}_{j}\right\rangle _{nn}=-0.3344(2)J$, and a staggered
magnetization, $m=0.296(2)$. These results are essentially the same as the
best variational result obtained by LDA, and also compare extremely well
with exact numerical results \cite{lda_88}, $\left\langle \mathbf{S}%
_{i}\cdot \mathbf{S}_{j}\right\rangle _{nn}=-0.3346$, and $m=0.31$.

Therefore, at half filling, the insulating state $|\mathrm{RVB}\rangle$
correctly reproduces both short range and long range AF correlations of the
Heisenberg model. AFLRO appears naturally in the long wavelength limit, with
a correct magnitude of magnetization; short range spin correlations are also
well described as evidenced by the excellent agreement of our results with
the ground state energy obtained from numerical studies.

\subsubsection{ Spin liquid at finite doping}

At finite doping, the RVB amplitudes $W_{ij}$, determined by (\ref{hs0}) and
(\ref{wij}) decay exponentially for large spatial separations,
\begin{equation}
\left\vert W_{ij}\right\vert \propto e^{-\frac{|\mathbf{r}_{ij}|^{2}}{2\xi
^{2}}}  \label{wij1}
\end{equation}%
with a characteristic length scale
\begin{equation}
\xi \simeq a\sqrt{\frac{2}{\pi \delta }}  \label{xi}
\end{equation}%
as shown in Fig. 3. Correspondingly, AFLRO disappears and a finite length
scale, $\xi $, for equal time spin correlations emerges; \textit{viz.}, a
featureless RVB spin liquid with short range correlations evolves
continuously from an AF ordered state at half filling ($\xi \rightarrow
\infty $).
\begin{figure}[tbph]
\includegraphics {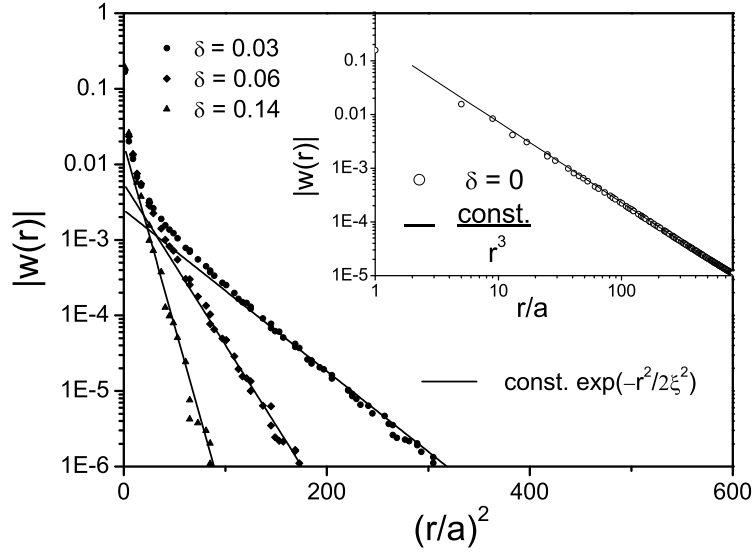}
\caption{The RVB amplitudes, $|W_{ij}|$, obtained from the b-RVB theory for
three different hole concentrations. The inset shows the result for half
filling.}
\end{figure}

Spin excitations can be created by the action of a spin raising/lowering
operator on the spin liquid $|\mathrm{RVB}\rangle $ in (\ref{rvb1}). In the
bosonic mean field theory, spin excitations are described by the spinon RVB
Hamiltonian, (\ref{hs0}). The results from mean field theory have been
reported elsewhere \cite{weng_99,chen_04}, and we briefly summarize the
results below. At small doping, the spin excitations evolve (from the
gapless magnon mode at half filling) into a sharp resonance at low energies,
with a characteristic energy scale, $E_{g}\propto \delta J$, around the
momentum $(\pi ,\pi ).$ The spectral weight of this excitation is
proportional to $\delta $, while its width in the momentum space is
proportional to $\xi ^{-1}$, which follows from (\ref{wij1}). The evolution
of the gapless magnon mode at half filling into a resonant mode at finite
energies reflects the phase string effect on the spin degrees of freedom,
through the effect of the gauge field, $A_{ij}^{h}$.

\subsection{Charge degrees of freedom: optimizing the kinetic energy of
doped holes}

At finite doping, the mean field state $|\mathrm{RVB}\rangle _{\mathrm{MF}}$
describes a condensate of RVB spinon pairs, and does not involve the charge
degrees of freedom (holons) explicitly; \textit{i.e.}, the distribution of
holons is independent of the positions of spinons. Since the single
occupancy constraint is no longer crucial in the bosonic RVB theory, it is
tempting to posit a mean field ground state
\begin{equation}
|\mathrm{GS}\rangle =|\Psi \rangle _{h}\otimes |\mathrm{RVB}\rangle _{%
\mathrm{MF}}  \label{GS}
\end{equation}%
where $|\Psi \rangle _{h}$ describes the holon condensate. This approach,
adopted in a previous work \cite{weng_99}, may be justified for a long
wavelength, low energy theory as the local constraint is not crucial in the
phase-string formalism \cite{remark1}. However, to obtain the ground state
wavefunction, it is desirable to employ a construction where the single
occupancy constraint is satisfied exactly; \textit{viz.}, a method that
treats spin backflow effects more carefully, which is also important for
facilitating and optimizing the kinetic energy of doped holes. In the
following, we show how this can be done differently.

First, we project out all double occupancies in $|\mathrm{RVB}\rangle _{%
\mathrm{MF}}$, and define $|\mathrm{RVB}\rangle $ as given in (\ref{rvb1}),
where each site is occupied only by one spinon, as if the system were at
half filling. However, the RVB amplitudes, $W_{ij}$, will be determined by
the bosonic RVB mean field theory of $H_{s}$ at a fixed hole concentration.

Next, for every doped holon at site $l$, a spinon at the same site $l$ is
annihilated, by the action of $h_{l}^{\dagger }b_{l\sigma }$ on $|\mathrm{RVB%
}\rangle $, \textit{i.e.},
\begin{equation}
\left( h_{l}^{\dagger }b_{l\sigma }\right) |\mathrm{RVB}\rangle  \label{h-s}
\end{equation}%
such that the constraint is automatically satisfied. As discussed below, the
construction (\ref{h-s}) optimizes the kinetic energy of the holon-spinon
composite, $h_{l}^{\dagger }b_{l\sigma }$, by maximizing the overlap between
the RVB configurations before and after hopping. Clearly, the behavior of
the spinon $b_{l\sigma }$ characterizes the spinon backflow accompanying the
motion of the holon in the RVB background.

\begin{figure}[tbph]
\includegraphics{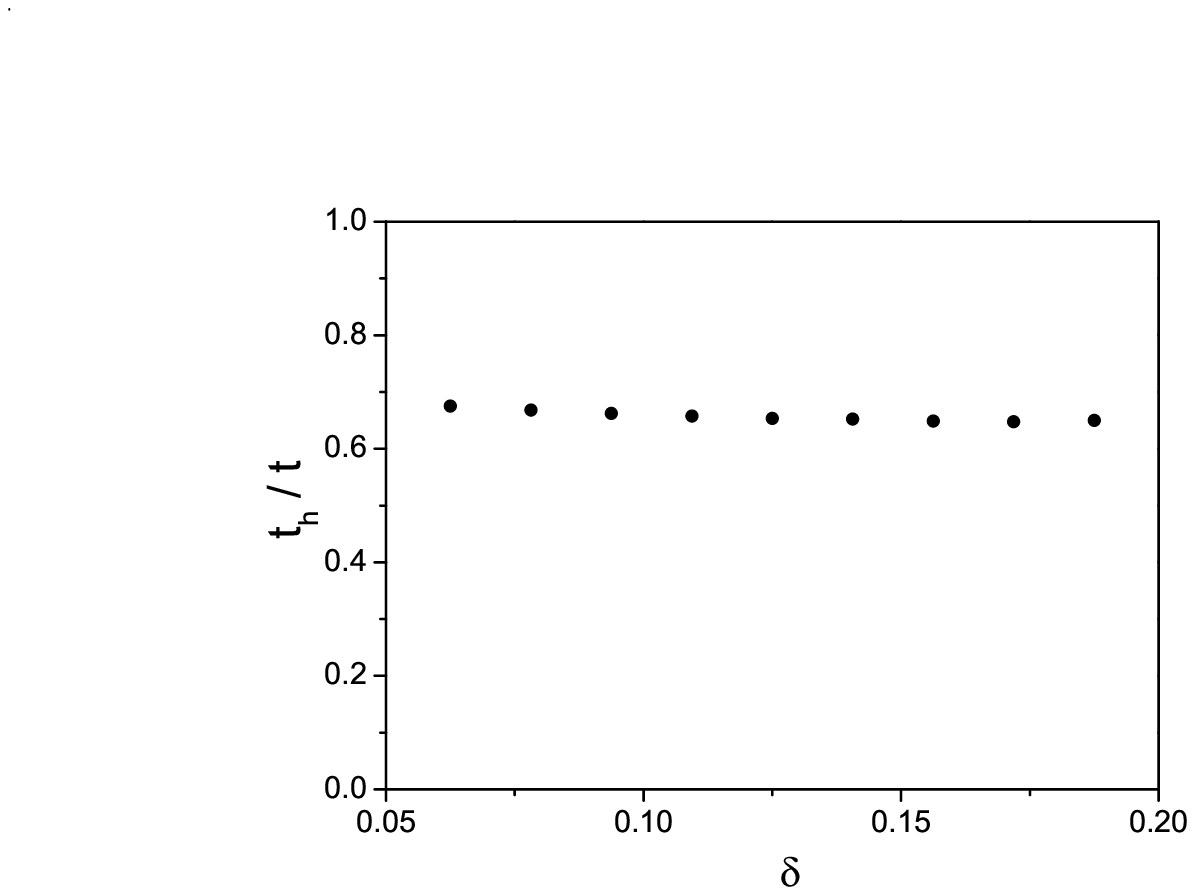}
\caption{Ratio of the effective hopping integral $\frac{t_h}{t}$ obtained
from (\protect\ref{zeq}).}
\end{figure}

Thus, the full ground state may be constructed as
\begin{equation}
|\Psi _{G}\rangle =\sum_{\{l_{h}\}}\varphi
_{h}(l_{1},l_{2,}...,l_{N_{h}})\prod_{h}Z_{\sigma _{h}}(l_{h})\left(
h_{l_{h}}^{\dagger }b_{l_{h}\sigma _{h}}\right) |\mathrm{RVB}\rangle ~,
\label{gsph0}
\end{equation}%
where $\varphi _{h}(\{l_{h}\})$ is a bosonic (symmetric) many body wave
function for holons, and $Z_{\sigma _{h}}(l_{h})$ denotes a single particle
wave function for the spinon backflow $b_{l_{h}\sigma _{h}}$, with the spin
index $\sigma _{h}$ under the constraint $1/2\sum_{h}\sigma _{h}=S_{z}$
(note that $S_{z}=0$ in $|\mathrm{RVB}\rangle )$.

In the following, we describe how to determine $\varphi _{h}(\{l_{h}\})$ and
$Z_{\sigma _{h}}(l_{h})$ in (\ref{gsph0}) based on a mean-field
approximation. For this purpose, we introduce a mean-field ground state as
follows
\begin{equation}
|\Psi _{G}\rangle \equiv P_{G}|\Psi _{G}\rangle _{\mathrm{MF}}
\end{equation}%
where the mean-field state
\begin{equation}
|\Psi _{G}\rangle _{\mathrm{MF}}=\sum_{\{l_{h}\}}\varphi
_{h}(l_{1},l_{2,}...,l_{N_{h}})\prod_{h}Z_{\sigma _{h}}(l_{h})\left(
h_{l_{h}}^{\dagger }b_{l_{h}\sigma _{h}}\right) |\mathrm{RVB}\rangle _{%
\mathrm{MF}}~  \label{gsph}
\end{equation}%
By optimizing the kinetic energy of the holon-spinon composite, $%
h_{l}^{\dagger }b_{l\sigma }$, based on (\ref{gsph}), $\varphi
_{h}(\{l_{h}\})$ and $Z_{\sigma _{h}}(l_{h})$ are determined as follows. As
shown in Appendix A, one finds
\begin{equation}
\varphi _{h}(l_{1},l_{2,}...,l_{N_{h}})=\text{\textrm{cons}}\mathrm{t.}
\label{hcond}
\end{equation}%
namely, the holons are in a Bose condensed state; $Z_{\sigma _{h}}(l_{h})$
satisfies the following eigen equation
\begin{equation}
(-t_{h})Z_{\sigma _{h}}(i)=-\frac{\tilde{t}}{4}\sum_{j=nn(i)}e^{-i\phi
_{ij}^{0}-i\sigma _{h}{A}_{ij}^{h}}Z_{\sigma _{h}}(j)~,  \label{zeq}
\end{equation}%
with $t_{h}>0$ being the maximal eigenvalue such that the total hopping
energy is given by
\begin{equation}
\langle \Psi _{G}|H_{t}|\Psi _{G}\rangle _{\mathrm{MF}}/\langle \Psi
_{G}|\Psi _{G}\rangle _{\mathrm{MF}}=-\ 4t_{h}\delta N~,
\end{equation}%
and the ``bare'' hopping integral
\begin{equation}
\tilde{t}=\left( \frac{\bar{n}^{b}}{2}+\frac{\left| \Delta ^{s}\right| ^{2}}{%
2\bar{n}^{b}}\right) t,\text{ \qquad }\bar{n}^{b}=1-\delta ~.  \label{th}
\end{equation}%
Note that $\Delta ^{s}\sim 1$ at small $\delta $, one has $\tilde{t}\sim t$
in terms of (\ref{th}). So the holes acquire a good ``bare'' hopping
integral in the bosonic RVB background since the singular frustrations due
to the phase string effect are gauged away in the phase string formalism
[one would get a minus sign in the second (RVB) term in (\ref{th}) in the
original slave-fermion representation]. The residual frustrations of the
phase string effect are represented by the phase field $\phi _{ij}^{0}$ and $%
{A}_{ij}^{h}$ in the eigen equation (\ref{zeq}) which determines the
renormalized hopping integral $t_{h}$. The latter is reduced from $\tilde{t}$
as shown in Fig. 4 numerically. In Appendix B, we briefly discuss how to go
beyond the above mean-field construction in an effective model description,
which is essentially the same as obtained in Ref.\onlinecite{weng_99}.

\section{Pseudogap and superconducting ground states in electron
representation}

So far, we discussed the reformulation of the $t-J$ model using the phase
string representation, where the singular effects arising from the
constraint on occupancy and the phase string, are sorted out explicitly.
This allowed us to develop an approximate theory in the above section. In
this section, we will rewrite our results in terms of the underlying
electronic degrees of freedom, and discuss the physical consequences and
implications.

\subsection{The ground state (\protect\ref{gsph0}) in electron representation%
}

It is straightforward to reexpress the ground state (\ref{gsph0}) in terms
of the electron operators based on (\ref{mutual})
\begin{eqnarray}
|\Psi _{G}\rangle &=&\sum_{\{l_{h}\}}\varphi
_{h}(l_{1},l_{2,}...,l_{N_{h}})\prod_{h}Z_{\sigma _{h}}(l_{h})\left( e^{-i%
\hat{\Theta}_{l_{h}\sigma _{h}}}~c_{l_{h}\sigma _{h}}\right) |\mathrm{RVB}%
\rangle  \nonumber \\
&=&\sum_{\{l_{h}\}}\varphi _{h}(l_{1},l_{2,}...,l_{N_{h}})\prod_{h}Z_{\sigma
_{h}}(l_{h})\tilde{c}_{l_{h}\sigma _{h}}|\mathrm{RVB}\rangle  \label{elgs0}
\end{eqnarray}%
where we introduce a new \textquotedblleft hole\textquotedblright\ creation
operator $\tilde{c}_{l\sigma }\equiv h_{i}^{\dagger }b_{i\sigma }$ which is
related to the electron operator by
\begin{equation}
\tilde{c}_{l\sigma }=e^{-i\hat{\Theta}_{i\sigma }}c_{l\sigma }.
\label{bhole}
\end{equation}%
The \textquotedblleft twisted\textquotedblright\ hole operator, $\tilde{c}%
_{l\sigma }$, satisfies bosonic statistics because of the phase factor, $%
e^{-i\hat{\Theta}_{i\sigma }}$ (see Sec. IIA). On using the holon
condensation condition (\ref{hcond}) and noting that $c_{l\sigma }^{2}\equiv
0$ and $c_{l\uparrow }c_{l\downarrow }|\mathrm{RVB}\rangle \equiv 0$, the
ground state can be rewritten in its final compact form
\begin{equation}
|\Psi _{G}\rangle =\mathrm{const.}\left( \sum_{l}Z_{\uparrow }(l)\tilde{c}%
_{l\uparrow }\right) ^{N_{h}^{\uparrow }}\left( \sum_{l}Z_{\downarrow }(l)%
\tilde{c}_{l\downarrow }\right) ^{N_{h}^{\downarrow }}|\mathrm{RVB}\rangle ~
\label{elgs}
\end{equation}%
where $N_{h}^{\uparrow }+N_{h}^{\downarrow }=N_{h}$ is the total number of
the holes.

We can further reexpress $|\mathrm{RVB}\rangle $ in the electron
representation after some straightforward manipulations. By using (\ref{rvb1}%
) and (\ref{mutual}), we find
\begin{equation}
|\mathrm{RVB}\rangle =\mathrm{const.}\sum_{\{\sigma _{s}\}}\Phi _{\mathrm{RVB%
}}(\sigma _{1},\sigma _{2},\cdot \cdot \cdot \sigma _{N})c_{1\sigma
_{1}}^{\dagger }c_{2\sigma _{2}}^{\dagger }\cdot \cdot \cdot c_{N\sigma
_{N}}^{\dagger }|0\rangle ~  \label{rvb3}
\end{equation}%
where the RVB wave function, $\Phi _{\mathrm{RVB}}$, is defined by
\begin{equation}
\Phi _{\mathrm{RVB}}(\sigma _{1},\sigma _{2},\cdot \cdot \cdot ,\sigma
_{N})\equiv \sum_{\mathrm{pair}}\prod_{(ij)}(-1)^{i}W_{ij}.  \label{phirvb-1}
\end{equation}%
Note that $|\mathrm{RVB}\rangle $ is always half-filled, involving spins on
the whole lattice which satisfy the no double occupancy constraint. Here
each spin configuration $\{\sigma _{s}\}=\sigma _{1},\sigma _{2},\cdot \cdot
\cdot ,\sigma _{N}$ is weighted by $\Phi _{\mathrm{RVB}}$ in a way that
spins are all paired up with the RVB amplitudes, $(-1)^{i}W_{ij}.$ The
latter depend upon doping as determined by the mean field theory described
in the previous section (see Fig. 3). The summation in $\Phi _{\mathrm{RVB}}$
runs over all possible pairing partitions for each configuration $\{\sigma
_{s}\}$.

The ground state $|\Psi _{G}\rangle $, reduces to $|\mathrm{RVB}\rangle $ at
half-filling, where it describes the antiferromagnetism extremely well as
discussed in Sec. III A1. Upon doping, $|\mathrm{RVB}\rangle $ may be
regarded as an \emph{insulating} RVB vacuum. From (\ref{elgs}), we see that
each doped hole is ``twisted'' into a bosonic object $\tilde{c}$, \emph{via}
the nonlocal phase operator $\hat{\Theta}_{i\sigma }$, in order to optimize
its kinetic energy in such an RVB vacuum. The form of the state $|\Psi
_{G}\rangle $ suggests a generalized ``spin-charge separation''; \emph{viz. }%
, the bosonic holes created by $\tilde{c}$, move freely in the background $|%
\mathrm{RVB}\rangle $; they carry the charge while the vacuum is described
by the pairing of neutral spins (spinons). These two degrees of freedom are
indeed decoupled in the mean-field sense. But this is different from the
typical spin-charge separation because the twisted holes are accompanied by
spins. Indeed, annihilating a spinon $\sigma $ at the site $l$ in the RVB
vacuum by $\tilde{c}_{l\sigma }$ will always create an unpaired spinon $%
-\sigma $ within a scale $\xi $ where the spins were originally RVB paired
(with no free spinons). So the state $|\Psi _{G}\rangle $ is composed both
of RVB paired spinons and twisted holes as the holon-spinon pairs as in (57)
below.

Since the spins at the hole sites $l$'s in the vacuum $|\mathrm{RVB}\rangle $
are automatically annihilated by $\tilde{c}_{l\sigma }$'s, the no double
occupancy constraint is exactly imposed in $|\Psi _{G}\rangle$ at finite
doping. As shown in Appendix C1, the ground state retains time reversal
symmetry, which may not be obvious from the form of $|\Psi _{G}\rangle $. It
also respects global spin rotational symmetry under the approximation that
the twisted holes are perfectly Bose condensed, as discussed in Appendix C2.

The structure of $|\Psi _{G}\rangle $ in the doped case looks rather simple
in terms of the twisted holes moving on the RVB vacuum. But it would look
more complicated when written in terms of the $c$-operators in (\ref{elgs})
because $\hat{\Theta}_{i\sigma }$ is highly nontrivial. Physically, the
phase string $\hat{\Theta}_{i\sigma }$ can be regarded as the nonlocal phase
shift induced by doped holes. In the one dimensional case, where it is
responsible for the Luttinger liquid behavior, the phase string can be
identified precisely with the phase shift obtained from the Bethe ansatz
solution for the large-$U$ Hubbard model \cite{weng_97}. In the two
dimensional case, the role of the phase string operator $\hat{\Theta}
_{i\sigma }$ may be seen more clearly in the wavefunction form discussed in
Sec. II, \emph{i.e.}, (\ref{wavefunction}). Below we rewrite the
wavefunction (\ref{elgs}) in terms of the underlying electrons.

Starting from the wavefunction $\psi _{b}$, corresponding to the ground
state (\ref{gsph0}) in the basis of (\ref{b-basis}), and using the relation
between the wavefunctions in the two basis sets (\ref{wavefunction}), we
get,
\begin{equation}
\psi _{e}(\{i_{u}\};\{j_{d}\})=\mathcal{K}\sum_{pair}\left[
\prod_{(ij)}(-1)^{i}W_{ij}\prod_{(l_{1}j_{1})}X_{l_{1}j_{1}}%
\prod_{(i_{1}l_{1}^{\prime })}\bar{X}_{i_{1}l_{1}^{\prime
}}\prod_{(ll^{\prime })}Y_{ll^{\prime }}\right]~.  \label{psie1}
\end{equation}%
Here, we have invoked the holon condensation condition (\ref{hcond}). In (%
\ref{elgs}), a given spin-hole configuration, $\{i_{u}\}$, $\{j_{d}\}$ and $%
\{l_{h}\}$, is partitioned into pairs, denoted by $(ij)$ (the sites, $i$ and
$j$, are occupied by an up and a down spin, respectively), $(l_{1}j_{1})$
(sites $l_{1}$ and $j_{1}$ are occupied by a hole and a down spin,
respectively), $(i_{1}l_{1}^{\prime })$ (two sites, $i_{1}$ and $%
l_{1}^{\prime }$, are occupied by an up spin and a hole, respectively), and $%
(ll^{\prime })$ (a pair of holes). The latter three pairings have
amplitudes, $X_{l_{1}j_{1}}\equiv Z_{\uparrow }(l_{1})W_{l_{1}j_{1}}$, $\bar{%
X}_{i_{1}l_{1}^{\prime }}\equiv (-1)^{i_{1}}W_{i_{1}l_{1}^{\prime
}}Z_{\downarrow }(l_{1}),$ and $Y_{ll^{\prime }}\equiv Z_{\uparrow
}(l)Z_{\downarrow }(l^{\prime })W_{ll^{\prime }}$, respectively. Fig. 5
shows two possible partitions for a given spin-hole configuration, and the
summation in (\ref{psie1}) runs over all possible partitions. Thus, the
wavefunction (\ref{psie1}) shows that the phase string operator $\hat{\Theta}%
_{i\sigma }$ leads to the generalized Jastrow-type factor $\mathcal{K}$
defined in (\ref{k2}). The remanining part of the wavefunction is bosonic,
describing RVB pairing of spins, pairing between spins and holes as well as
between holes.

\begin{figure}[tbph]
\includegraphics[width=2.5in]{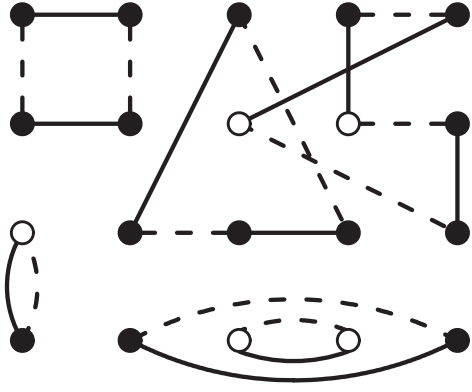}
\caption{Pairing among spins (full circles) and holes (open circles) are
illustrated by the connection of lines. Two sets of bonds (solid and dashed
lines) represent two different resonating pairing patterns, forming a loop
covering of the lattice for a given spin-hole configuration.}
\end{figure}

\subsection{Hidden ODLRO in the ground state: A pseudogap state}

The properties of the RVB vacuum $|\mathrm{RVB}\rangle $ were discussed in
Sec. III A. In the following we shall focus on the charge degrees of freedom
in $|\Psi _{G}\rangle $.

>From the definition of (\ref{elgs}), it follows that
\begin{equation}
\langle \Psi _{G}|\left( \sum_{l}Z_{\sigma }(l)\tilde{c}_{l\sigma }\right)
|\Psi _{G}\rangle \simeq \sqrt{\delta N}  \label{odlro1}
\end{equation}%
for finite doping: $N_{h}^{\uparrow ,\downarrow }=O(\delta N/2)$. Thus the
ground state exhibits an ODLRO as the twisted holes, described by $\tilde{c}%
_{i\sigma }$ [or more precisely the bosonic object, $\sum_{l}Z_{\sigma }(l)%
\tilde{c}_{l\sigma }$, which is gauge-invariant under (\ref{u(1)1}) and (\ref%
{u(1)2})], are Bose condensed on the RVB vacuum $|\mathrm{RVB}\rangle $ in (%
\ref{elgs}). Since $\tilde{c}_{i\sigma }$ is a nonlocal operator which
differs from the electron operator by an infinite-body phase-string twist,
the corresponding ODLRO in (\ref{odlro1}) is unlike the conventional ODLRO
based on a local combination of electron operators, such as spin
magnetization or the BCS superconducting pairing operator. In other words,
it does not have a direct correspondence to a physical observable. Such a
hidden ODLRO characterizes how doped charges (holes) are collectively
reorganized and behave in the RVB background, $|\mathrm{RVB}\rangle $. As
the consequences, the ground state and excited states will exhibit some very
unique novel properties. This will be the subject of our discussion below.

\subsubsection{Pairing amplitude}

We first show that the hidden ODLRO in $|\Psi _{G}\rangle $ will result in a
finite pairing amplitude between electrons in the singlet channel.

The electron singlet pair operator can be expressed as
\begin{eqnarray}
\hat{\Delta}_{ij}^{\mathrm{SC}} &\equiv &\sum_{\sigma }\sigma c_{i\sigma
}c_{j-\sigma }\   \nonumber \\
&=&\sum_{\sigma }\sigma e^{i\hat{\Theta}_{i\sigma }}\text{ }\tilde{c}%
_{i\sigma }e^{i\hat{\Theta}_{j-\sigma }}\tilde{c}_{j-\sigma }\   \nonumber \\
&=&\hat{\Delta}_{ij}^{0}e^{i\frac{1}{2}\left( \Phi _{i}^{s}+\Phi
_{j}^{s}\right) }~.  \label{sco}
\end{eqnarray}%
In the last line, the pair operator is formally written in terms of an
amplitude and a phase part, with the amplitude operator defined by \cite%
{zhou_03}
\begin{equation}
\hat{\Delta}_{ij}^{0}\equiv \sum_{\sigma }\left[ (-1)^{j}e^{-i\Phi
_{j}^{0}-i\phi _{ij}^{0}-i\sigma A_{ij}^{h}}\right] \tilde{c}_{i\sigma }%
\tilde{c}_{j-\sigma }~  \label{sca}
\end{equation}%
(a constant factor $(-1)^{N_{h}}$ is omitted).

We now demonstrate that $\left\langle \hat{\Delta}_{ij}^{0}\right\rangle
\neq 0$, \textit{i.e.}, the pairing amplitude is finite as the consequence
of the hidden ODLRO. Since the twisted holes are Bose condensed, $\left(
\tilde{c}_{i\sigma }\tilde{c}_{j-\sigma }\right) ^{\dagger }|\Psi
_{G}\rangle _{N_{e-2}}=A|\Psi _{G}\rangle _{N_{e}}+\mathrm{excited\ states}$%
, where the coefficient $A$ is nonzero and can be estimated by $A\sim
Z_{\sigma }(i)Z_{-\sigma }(j)\sqrt{N_{h}^{\uparrow }N_{h}^{\downarrow }}$ at
small doping \cite{remark3}. The pair amplitude $\Delta
_{ij}^{0}=_{N_{e}-2}\langle \Psi _{G}|\hat{\Delta}_{ij}^{0}|\Psi _{G}\rangle
_{N_{e}}$ is then given by
\begin{eqnarray}
\Delta _{ij}^{0} &\simeq &\sqrt{N_{h}^{\uparrow }N_{h}^{\downarrow }}%
\sum_{\sigma }(-1)^{j}e^{-i\Phi _{j}^{0}-i\phi _{ij}^{0}-i\sigma
A_{ij}^{h}}Z_{\sigma }^{\ast }(i)Z_{-\sigma }^{\ast }(j)\   \nonumber \\
&=&\frac{\delta N}{2}\sum_{\sigma }e^{-i\phi _{ij}^{0}-i\sigma
A_{ij}^{h}}Z_{\sigma }^{\ast }(i)Z_{\sigma }(j)~,  \label{d01}
\end{eqnarray}%
where we have used $N_{h}^{\uparrow }\sim N_{h}^{\downarrow }\sim
N_{h}/2=\delta N/2$, and
\begin{equation}
Z_{-\sigma }^{\ast }(j)=Z_{\sigma }(j)(-1)^{j}e^{i\Phi _{j}^{0}}~.
\label{zd}
\end{equation}%
The above equation can be obtained from (\ref{zeq}), and by noting that $%
e^{i\Phi _{i}^{0}-i\Phi _{j}^{0}}=-e^{i2\phi _{ij}^{0}}$. In order to
estimate the size of the pair amplitude, we focus on the pairing amplitude
between nn sites, $(\Delta _{ij}^{0})_{nn}$. Invoking translational
invariance of the ground state, we get
\begin{eqnarray}
\Delta ^{0} &\equiv &\frac{1}{2N}\sum_{<ij>}\left( \Delta _{ij}^{0}\right)
_{nn}\   \nonumber \\
&=&\frac{\delta }{4}\sum_{<ij>}\sum_{\sigma }e^{-i\phi _{ij}^{0}-i\sigma
A_{ij}^{h}}Z_{\sigma }^{\ast }(i)Z_{\sigma }(j)\   \nonumber \\
&=&\frac{\delta }{4}\sum_{\sigma }\frac{2t_{h}}{\tilde{t}}\sum_{i}\left|
Z_{\sigma }(i)\right| ^{2}\   \nonumber \\
&=&\frac{t_{h}}{\tilde{t}}\delta ~~~.  \label{d0}
\end{eqnarray}%
Therefore the ground state (\ref{elgs}) indeed possesses a finite pairing
amplitude as a direct consequence of the hidden ODLRO in $|\Psi _{G}\rangle $%
, which is proportional to the doping concentration for small doping. Note
that a finite pairing amplitude exists only in the singlet channel. It is
easily shown that the mean value of the pairing amplitude in the triplet
channel vanishes identically [in this case, an additional sign $\sigma $
appears inside the summation of (\ref{sca})].

However, superconducting phase coherence is generally absent. The phase part
of the pair operator (\ref{sco}) involves $\Phi _{i}^{s}$ which is defined
in (\ref{phis}) and is closely related to the spin degrees of freedom. More
precisely, each unpaired spin contributes to a free $\pm 2\pi $ vortex via $%
\Phi _{i}^{s}$ in the pair operator (\ref{sco}). As will be discussed in
subsection C, the presence of free spins in the ground state $|\Psi
_{G}\rangle $ prevents phase coherence in (\ref{sco}). Nevertheless, $%
\left\langle \hat{\Delta}_{ij}^{0}\right\rangle \neq 0$ in (\ref{sco}) is a
meaningful definition and description of a low temperature \emph{pseudogap}
ground state, called the spontaneous vortex phase \cite{weng_02}. In this
phase, each unpaired neutral spin (spinon) always carries a $\pm 2\pi $
supercurrent vortex, known as a \emph{spinon-vortex} \cite{vnm_02}. We
postpone further discussion of the spinon-vortices to Sec. IV C.

\subsubsection{Fermionic nodal quasiparticles}

In the following, we construct a quasiparticle excitation based on the
ground state $|\Psi _{G}\rangle $. In particular, we show that the existence
of nodes in the excitation spectrum is ensured by the hidden ODLRO.

Consider the action of $c^{\dagger }$ on the ground state $|\Psi _{G}\rangle
$. Since $c_{i\sigma }=\tilde{c}_{i\sigma }e^{i\hat{\Theta}_{i\sigma }}$ and
the twisted bosonic holes $\tilde{c}_{i\sigma }$ are condensed, one finds
\begin{eqnarray}
c_{i\sigma }^{\dagger }|\Psi _{G}\rangle _{N_{e}-1} &=&e^{-i\hat{\Theta}%
_{i\sigma }}\tilde{c}_{i\sigma }^{\dagger }|\Psi _{G}\rangle _{N_{e}-1}\
\nonumber \\
&\simeq &e^{-i\frac{1}{2}\Phi _{i}^{s}}B_{i\sigma }|\Psi _{G}\rangle
_{N_{e}}+\text{\textrm{excited states}}~,  \label{qp1}
\end{eqnarray}%
where
\begin{eqnarray}
B_{i\sigma } &\simeq &\sqrt{N_{h}^{\sigma }+1}\left( Z_{\sigma }(i)e^{-i\hat{%
\Theta}_{i\sigma }}e^{i\frac{1}{2}\Phi _{i}^{s}}\right) \   \nonumber \\
&\simeq &\sqrt{\frac{\delta N}{2}}(\sigma )^{N_{h}}(-\sigma )^{i}Z_{\sigma
}(i)e^{i\frac{1+\sigma }{2}\Phi _{i}^{0}}e^{-i\frac{\sigma }{2}\left( \Phi
_{i}^{0}-\Phi _{i}^{h}\right) }~.  \label{B}
\end{eqnarray}

Let us focus on $B_{i\sigma }$ first. Rewriting $B_{i\sigma }$ using (\ref%
{zd}), we get
\begin{equation}
B_{i\uparrow }=\sqrt{\frac{\delta N}{2}}Z_{\downarrow }^{\ast }(i)e^{i\frac{1%
}{2}\left( \Phi _{i}^{0}-\Phi _{i}^{h}\right) }~,
\end{equation}%
and
\begin{equation}
B_{i\downarrow }=\sqrt{\frac{\delta N}{2}}(-1)^{N_{h}}Z_{\downarrow }(i)e^{-i%
\frac{1}{2}\left( \Phi _{i}^{0}-\Phi _{i}^{h}\right) }~.
\end{equation}%
Nontrivial spatial oscillations in $B_{i\sigma }$ mainly originate from the
phase factor $e^{-i\frac{\sigma }{2}\left( \Phi _{i}^{0}-\Phi
_{i}^{h}\right) }$. We rewrite this phase factor as a string operator,
\begin{equation}
e^{-i\frac{\sigma }{2}\left[ \Phi _{i}^{0}-\Phi _{i}^{h}\right] }=e^{-i\frac{%
\sigma }{2}\left[ \Phi _{i}^{0}-\Phi _{i}^{h}\right] }e^{i\frac{\sigma }{2}%
\left[ \Phi _{i_{1}}^{0}-\Phi _{i_{1}}^{h}\right] }e^{-i\frac{\sigma }{2}%
\left[ \Phi _{i_{1}}^{0}-\Phi _{i_{1}}^{h}\right] }e^{i\frac{\sigma }{2}%
\left[ \Phi _{i_{2}}^{0}-\Phi _{i_{2}}^{h}\right] }\cdot \cdot \cdot ~,
\label{stringqp1}
\end{equation}%
where $i,$ $i_{1},$ $i_{2,}\cdot \cdot \cdot $ is a sequence of nn sites on
a path ending at site $i$. Using the relation $\theta _{i}(i_{1})-\theta
_{i_{1}}(i)=\pm \pi $, and invoking the Bose condensation of the holes, we
get
\begin{equation}
e^{-i\frac{\sigma }{2}\left[ \Phi _{i}^{0}-\Phi _{i}^{h}\right] }e^{i\frac{%
\sigma }{2}\left[ \Phi _{i_{1}}^{0}-\Phi _{i_{1}}^{h}\right] }\simeq e^{\mp
i\sigma \frac{\pi }{2}(1-\eta )}e^{-i\sigma \left[ \phi
_{ii_{1}}^{0}-A_{ii_{1}}^{h}\right] },\text{ \quad }etc.,  \label{stringqp}
\end{equation}%
with $\eta =\delta $ such that
\begin{equation}
B_{i\sigma }B_{j\sigma }^{\dagger }=D_{ij}^{\sigma }e^{-i\sigma \mathbf{k}%
_{0}\cdot (\mathbf{r}_{i}-\mathbf{r}_{j})}
\end{equation}%
where
\begin{equation}
\mathbf{k}_{0}=\left( \pm \frac{\pi }{2}[1-\eta ],\pm \frac{\pi }{2}[1-\eta
]\right)  \label{k01}
\end{equation}%
at four points in the Brillouin zone (setting the lattice constant $a=1$).
The quasiparticle weight is proportional to
\begin{eqnarray}
D_{ij}^{\uparrow } &=&\left( D_{ij}^{\downarrow }\right) ^{\ast }\
\nonumber \\
&=&\frac{\delta N}{2}Z_{\downarrow }^{\ast }(i)e^{-i\sum_{\Gamma _{ij}}\left[
\phi _{lm}^{0}-A_{lm}^{h}\right] }Z_{\downarrow }(j)\   \nonumber \\
&\sim &O(\delta )~,  \label{D}
\end{eqnarray}%
which is gauge invariant, as can be seen from (\ref{zeq}). Here, $\Gamma
_{ij}$ denotes a shortest-path connecting sites $i$ and $j$.

Therefore one finds
\begin{equation}
c_{i\sigma }^{\dagger }|\Psi _{G}\rangle _{N_{e}-1}\simeq e^{-i\sigma
\mathbf{k}_{0}\cdot \mathbf{r}_{i}}e^{-i\frac{1}{2}\Phi _{i}^{s}}|\Psi
_{G}\rangle _{N_{e}}.
\end{equation}%
If phase coherence is realized such that
\begin{equation}
e^{-i\frac{1}{2}\Phi _{i}^{s}}|\Psi _{G}\rangle \sim |\Psi _{G}\rangle
\label{pcoherence}
\end{equation}%
(see Sec. IV C), then $c_{\mathbf{k}_{0}\sigma }^{\dagger }|\Psi _{G}\rangle
_{N_{e}-1}$ would have a finite overlap with $|\Psi _{G}\rangle _{N_{e}}$;
namely, the quasiparticle spectrum has four nodes at momenta, $\mathbf{k}%
_{0} $'s, that approach $\left( \pm \frac{\pi }{2},\pm \frac{\pi }{2}\right)
$ in the limit of zero doping. Similar momentum structure can be identified
in $c_{\mathbf{k}\sigma }|\Psi _{G}\rangle _{N_{e}+1}$. A true d-wave-like
quasiparticle state in the superconducting state can be constructed as a
linear combination of $c_{\mathbf{k}\sigma }^{\dagger }|\Psi _{G}\rangle $
and $c_{-\mathbf{k-}\sigma }|\Psi _{G}\rangle $, which should be
infinitesimally close to the ground state $|\Psi _{G}\rangle _{N_{e}}$ at
the four nodal points, $\mathbf{k}_{0}$'s.

However, as discussed in Sec. IV C, the phase factor $e^{-i\frac{1}{2}\Phi
_{i}^{s}}$ precludes phase coherence, and therefore the quasiparticle state $%
c_{i\sigma }^{\dagger }|\Psi _{G}\rangle _{N_{e-1}}$ actually has a
vanishing overlap with the ground state $|\Psi _{G}\rangle _{N_{e}}$ because
of such a phase factor. Thus, the low-lying fermionic quasiparticles near
the four nodes, $\mathbf{k}_{0}$'s, have a vanishing spectral weight $Z_{%
\mathbf{k}}\equiv |_{N_{e}}\langle \Psi _{G}|c_{\mathbf{k}\sigma }^{\dagger
}|\Psi _{G}\rangle _{N_{e}-1}|^{2}=0$; \textit{viz.}, the
non-superconducting (pseudogap) ground state $|\Psi _{G}\rangle $ is a
non-Fermi-liquid in nature. Long lived quasiparticle excitations are
expected to emerge only when superconducting phase coherence is realized in (%
\ref{pcoherence}).

We emphasize that the nodal structure of quasiparticles results from the
ODLRO of the bosonic twisted holes, in which the phase string factor $e^{-i%
\hat{\Theta}_{i\sigma }}$, reflecting the fermionic nature of the electron
operator, determines the momentum positions of four nodes. Finally we note
that in the previous effective phase string theory in terms of holons and
spinons (see Appendix B), a similar d-wave quasiparticle excitation can be
obtained as the consequence of forming the holon-spinon bound pairs by
considering residual interactions beyond the effective model \cite{zhou_03}.

\subsubsection{Two types of spin excitations}

The hidden ODLRO in (\ref{elgs}) is not only responsible for the existence
of a finite pairing amplitude as well as d-wave like nodal quasiparticle
excitations, it also predicts the presence of a new type of low-lying spin
excitations which are quite distinctive as compared to the previously
discussed spin excitations created in the RVB background $|\mathrm{RVB}%
\rangle $ (Sec. III A)$.$

The latter may be called Type I spin excitations, which can be created by
acting a spin operator, say, $S_{i}^{+}$, directly onto the insulating spin
liquid $|\mathrm{RVB}\rangle $, without involving the \emph{hole} condensate
in (\ref{elgs}). Obviously, this type of spin excitations is the only
elementary excitation at half-filling, where it reduces to a conventional
spin wave in the long-wavelength limit as discussed in Sec. III A. At small
doping, the majority of spins still remain RVB paired, characterized by $%
\Delta ^{s}\neq 0$ in $|\mathrm{RVB}\rangle .$\ Thus, this kind of spin
excitation will naturally persist into the pseudogap regime, where a \emph{%
spin} \emph{gap }will be opened up in its low-energy spectrum as the
spin-spin correlations are short-ranged (here $W_{ij}$ becomes exponential
decay with the distance $\left| i-j\right| ,$ see Fig. 3). In fact, as
mentioned in Sec. III A, the corresponding low-lying excitation reduces to a
sharp resonance-like mode in the pseudogap regime, with a characteristic
energy scale $E_{g}\sim \delta J$ around the momentum $(\pi ,\pi )$. Remnant
spin wave like dispersion is still present at high energies \cite{chen_04},
which disappears when the temperature reaches $T_{0}$ as shown in Fig. 2
where the short-range RVB pairing vanishes finally.

Now let us discuss a new type (Type II) of $S=1$ excitation which is
associated with the twisted holes and may be regarded as composed of a pair
of \emph{quasiparticle and quasihole,} instead of a pair of neutral spinons
in $|\mathrm{RVB}\rangle $. To create a Type II spin excitation by, say, $%
S_{i}^{+},$ a pair of quasiparticles will be created and annihilated in $%
|\Psi _{G}\rangle $, which involves a spin flip in the ``hole condensate''
part in (\ref{elgs}).

To show this, let us rewrite $S_{i}^{+}=c_{i\uparrow }^{\dagger
}c_{i\downarrow }$ by
\begin{eqnarray}
S_{i}^{+} &=&e^{i\hat{\Theta}_{i\uparrow }}\tilde{c}_{i\uparrow }^{\dagger }%
\tilde{c}_{i\downarrow }e^{-i\hat{\Theta}_{i\downarrow }}  \nonumber \\
&=&\left[ (-1)^{i}e^{i\Phi _{i}^{h}}\right] \tilde{c}_{i\uparrow }^{\dagger }%
\tilde{c}_{i\downarrow }.  \label{s+1}
\end{eqnarray}%
Then, by using $\left( \cdot \cdot \cdot \right) _{\mathrm{hole-}\text{%
\textrm{part}}}$ to denote Type II spin excitations created by $S_{i}^{+}$
in the condensed ``hole'' part in $|\Psi _{G}\rangle $, we obtain
\begin{equation}
\left( S_{i}^{+}|\Psi _{G}\rangle \right) _{\mathrm{hole-}\text{\textrm{part}%
}}\simeq (-1)^{i}e^{i\Phi _{i}^{h}}Z_{\uparrow }(i)Z_{\downarrow }^{\ast }(i)%
\sqrt{N_{h}^{\uparrow }}\sqrt{N_{h}^{\downarrow }+1}|\Psi _{G}\rangle ~.
\label{typeii}
\end{equation}%
The term on the right-hand side (rhs) has the same origin as the
quasiparticles around nodal points discussed previously. But the phase
factor $e^{-i\hat{\Theta}_{i\sigma }}$ in the quasiparticle channel [(\ref%
{qp1})] is now replaced by $(-1)^{i}e^{i\Phi _{i}^{h}}$ in the spin channel,
after a recombination of $e^{i\hat{\Theta}_{i\uparrow }}$ and $e^{-i\hat{%
\Theta}_{i\downarrow }}$ in (\ref{s+1}) \cite{remark4}$.$ The leading
contribution to the spin-spin correlation at large distance is then given by
\begin{equation}
\langle \Psi _{G}|S_{j}^{-}S_{i}^{+}|\Psi _{G}\rangle _{\mathrm{hole-}\text{%
\textrm{part}}}\simeq (-1)^{i-j}\left\langle Z_{\uparrow }^{\ast
}(j)Z_{\downarrow }(j){\Pi }_{ij}^{h}Z_{\uparrow }(i)Z_{\downarrow }^{\ast
}(i)\right\rangle \delta N  \label{scorr}
\end{equation}%
(using $N_{h}^{\sigma }\sim \sqrt{\delta N})$ where
\begin{equation}
{\Pi }_{ij}^{h}\equiv e^{i\left( \Phi _{i}^{h}-\Phi _{j}^{h}\right) }.
\label{piphase}
\end{equation}

The momentum structure can be determined by the Fourier transformation of (%
\ref{scorr}), in a way similar to that in the quasiparticle channel. Note
that, an additional topological effect in $e^{i\Phi _{i}^{h}}$ (not in $e^{i%
\frac{1}{2}\Phi _{i}^{h}}$) requires a little more careful treatment. As
detailed elsewhere \cite{inc}, $(-1)^{i}e^{i\Phi _{i}^{h}}$ will lead to the
following incommensurate momenta, to leading order of approximation,
\begin{equation}
\mathbf{Q}_{\delta }\mathbf{=}\left[ \pm \pi ,\pm \pi (1-\kappa _{\delta })%
\right] \text{ \quad \textrm{and \quad }}\left[ \pm \pi (1-\kappa _{\delta
}),\pm \pi \right]  \label{inc}
\end{equation}%
with $\kappa _{\delta }\simeq 2\delta $.

On the other hand, the dynamics of (\ref{scorr}) is mainly decided by the
vortices in $e^{i\Phi _{i}^{h}}$ which is in turn determined by the charge
density. To leading order approximation, no superexchange interaction in the
spin degrees of freedom influences (\ref{scorr}). If one neglects the
fluctuation in $e^{i\Phi _{i}^{h}},$ the term on the rhs of (\ref{typeii})
would have a finite overlap with $|\Psi _{G}\rangle $ at the incommensurate
momenta $\mathbf{Q}_{\delta }$, which implies a zero mode or the presence of
an incommensurate AFLRO. But the true long-range order will be prevented by
the density fluctuations, because any small fluctuation in the charge
density will be accumulated in the phase factor $e^{i\Phi _{i}^{h}}$ in a
sufficiently long time and distance$,$ which in turn will result in a
slow-fluctuating incommensurate AF ordering above $|\Psi _{G}\rangle $,
instead of a true static incommensurate AFLRO. This low-lying spin
fluctuation can provide an alternative explanation, without invoking a
dynamic stripe picture, of low-energy results in neutron scattering
experiments on the cuprate superconductors.

Therefore, Type II spin excitations in (\ref{typeii}) are quite distinct
from Type I in nature. As a direct consequence of the hidden ODLRO, they are
absent at half-filling [the weight of this term is proportional to the
density of the charge condensate, from (\ref{typeii})]. One should thus
properly distinguish these two in a neutron scattering measurement, as a
unique prediction of this theory.

\subsection{Superconducting phase coherence}

As described above, the hidden ODLRO in $|\Psi _{G}\rangle $ leads to a
nonzero pairing amplitude. But it does not necessarily imply
superconductivity. Consider the action of the superconducting pairing
operator on the $N_{e}$ electron ground state. The resulting state has an
overlap with $N_{e}-2$ electron ground state, determined by using (\ref{sco}%
) and (\ref{d0}), as follows:
\begin{equation}
\left( \hat{\Delta}_{ij}^{\mathrm{SC}}\right) _{nn}|\Psi _{G}\rangle
_{N_{e}}\simeq \Delta ^{0}e^{-i\frac{1}{2}\left( \Phi _{i}^{s}+\Phi
_{j}^{s}\right) }|\Psi _{G}\rangle _{N_{e}-2}+\text{\textrm{excited states}}.
\label{dsc}
\end{equation}%
So superconductivity is established only when
\begin{equation}
\left\langle e^{-i\frac{1}{2}\left( \Phi _{i}^{s}+\Phi _{j}^{s}\right)
}\right\rangle \neq 0~,  \label{pc}
\end{equation}%
\textit{viz.}, the onset of superconductivity is determined by phase
coherence.

\subsubsection{Spinon vortices}

We noted earlier that each isolated ($S=1/2$) spin in $|\Psi _{G}\rangle $
means the creation of a $\pm 2\pi $ vortex in the phase of the
superconducting order parameter $\hat{\Delta}_{ij}^{\mathrm{SC}}$, via $\Phi
_{i}^{s}$, according to (\ref{dsc}). Thus, each spin automatically carries a
phase or supercurrent vortex, called a spinon-vortex, in the state described
by $|\Psi _{G}\rangle $. This was first identified in previous work based on
the effective theory \cite{vnm_02,weng_02}.

The $\pm 2\pi $ vortices centered at $\uparrow $ and $\downarrow $ spins in $%
|\mathrm{RVB}\rangle $ are paired as vortex-antivortex pairs, since the
underlying spins are always paired in the singlet RVB vacuum. Then, one has
\begin{equation}
\langle \mathrm{RVB}|e^{-i\frac{1}{2}\left( \Phi _{i}^{s}+\Phi
_{j}^{s}\right) }|\mathrm{RVB}\rangle \neq 0.  \label{pc1}
\end{equation}%
This is in analogy with the two-dimensional XY model below the
Kosterlitz-Thouless transition where vortex-antivortex pairs are bound
together. In our case, vortices and antivortices are associated with $%
\uparrow $ and $\downarrow $ spins, respectively. The condition that $%
S^{z}=0 $ in the singlet background, $|\mathrm{RVB}\rangle $, is equivalent
to the charge neutrality condition in the XY model. Since $\uparrow $ and $%
\downarrow $ spins are RVB pairs with a \emph{finite} range $\xi $ at finite
doping [see (\ref{xi})], the cancellation in $\frac{1}{2}\left( \Phi
_{i}^{s}+\Phi _{j}^{s}\right) $ is ensured at large distances. Thus, we see
that the spin RVB pairing in $|\mathrm{RVB}\rangle $ is very essential for
phase coherence to occur. Note that this condition fails in the AF phase,
where $\xi \rightarrow \infty $, or beyond the doping concentration $x_{%
\mathrm{RVB}}$ where the RVB pairing disappears.

However, the existence of free spins, associated with doped holes created by
$\tilde{c}_{l\sigma }$, will imply the presence of \emph{free} vortices in $%
|\Psi _{G}\rangle $. To see this, following (\ref{elgs}) and (\ref{pc1}),
one can write
\begin{eqnarray}
e^{-i\frac{1}{2}\left( \Phi _{i}^{s}+\Phi _{j}^{s}\right) }|\Psi _{G}\rangle
&=&e^{i\frac{1}{2}\left( \Lambda _{i}^{h}+\Lambda _{j}^{h}\right) }\left(
\sum_{l}Z_{\uparrow }(l)\tilde{c}_{l\uparrow }\right) ^{N_{h}^{\uparrow
}}\left( \sum_{l}Z_{\downarrow }(l)\tilde{c}_{l\downarrow }\right)
^{N_{h}^{\downarrow }}e^{-i\frac{1}{2}\left( \Phi _{i}^{s}+\Phi
_{j}^{s}\right) }|\mathrm{RVB}\rangle \   \nonumber \\
&\sim &e^{i\frac{1}{2}\left( \Lambda _{i}^{h}+\Lambda _{j}^{h}\right) }|\Psi
_{G}\rangle  \label{pc2}
\end{eqnarray}%
where $\Lambda _{i}^{h}\equiv \sum_{l\neq i}\theta _{i}(l)\left( \tilde{n}%
_{l\uparrow }^{h}-\tilde{n}_{l\downarrow }^{h}\right) $ with $\tilde{n}%
_{l\sigma }^{h}$ denoting the number operator of twisted holes of spin $%
\sigma $ at site $l$. Then, the Bose condensation of these free vortices in $%
|\Psi _{G}\rangle $ will disorder the phase of the superconducting pairing
order parameter according to (\ref{pc2}) (although $\Lambda _{i}^{h}$ itself
vanishes on average, its fluctuations will satisfy an area law in the Bose
condensed case).

Therefore the ground state $|\Psi _{G}\rangle $ is not a true
superconducting ground state due to the lack of phase coherence, although it
possesses a pairing amplitude as well as (incoherent) nodal quasiparticles.
It is characterized by the presence of unpaired spinon-vortices as a direct
consequence of the hidden ODLRO in $|\Psi _{G}\rangle $. Nontrivial Nernst
effect contributed by such unpaired spinon-vortices has been investigated
based on the effective phase string theory elsewhere \cite{weng_02}.

\subsubsection{Superconducting ground state}

The superconducting instability will take place when those unpaired spins
associated with twisted holes also become RVB paired. It reduces the
superexchange energy when those backflow spins form the RVB pairs, and most
importantly it removes the logarithmic-divergent energy associated with free
vortices as vortices-antivortices are paired up at low temperature.
Consequently the phase $\Lambda _{i}^{h}$ is cancelled out in (\ref{pc2})
and the true phase coherence (\ref{pc}) of the superconducting order
parameter can be finally established.

Corresponding to the pairing of those free spins associated with twisted
holes, the twisted holes themselves are also effectively paired and the
ground state should evolve from (\ref{elgs}) into the following form%
\begin{equation}
|\Psi _{G}\rangle _{SC}=\mathrm{const.}\left( \hat{D}\right) ^{\frac{N_{h}}{2%
}}|\mathrm{RVB}\rangle ~  \label{scgs}
\end{equation}%
in which
\begin{equation}
\hat{D}=\sum_{ij}g(i-j)\left[ \sum_{\sigma }Z_{\sigma }(i)Z_{-\sigma }(j)%
\tilde{c}_{i\sigma }\tilde{c}_{j-\sigma }\right]  \label{dd}
\end{equation}%
with $g(i-j)$ characterizing the pairing of the twisted holes. Note that in
the limit of no pairing, with $g(i-j)=$\textrm{\ constant, }equation\textrm{%
\ }(\ref{scgs}) will recover the pseudogap (spontaneous vortex phase) ground
state (\ref{elgs}) at $N_{h}^{\uparrow }=N_{h}^{\downarrow }=N_{h}/2$.

According to (\ref{scgs}), one has $\left\langle \hat{D}\right\rangle \neq 0$%
. By using (\ref{sco}) - (\ref{zd}), $\hat{D}$ can be further written as
follows
\begin{eqnarray}
\hat{D} &\simeq &\sum_{ij}g(i-j)\left[ \frac{2}{\delta N}\sum_{\sigma
}e^{-i\phi _{ij}^{0}-i\sigma A_{ij}^{h}}Z_{\sigma }^{\ast }(i)Z_{\sigma }(j)%
\frac{\hat{\Delta}_{ij}^{0}}{2}\right]  \nonumber \\
&=&\frac{1}{2}\sum_{ij}g(i-j)\Delta _{ij}^{0}\hat{\Delta}_{ij}^{0}  \nonumber
\\
&=&\frac{1}{2}\sum_{ij}g(i-j)\Delta _{ij}^{0}\left[ e^{-i\frac{1}{2}\left(
\Phi _{i}^{s}+\Phi _{j}^{s}\right) }\hat{\Delta}_{ij}^{\mathrm{SC}}\right]
\label{dd1}
\end{eqnarray}%
such that

\begin{equation}
\left\langle \hat{D}\right\rangle =\frac{1}{2}\sum_{ij}g(i-j)\Delta
_{ij}^{0}\left\langle e^{-i\frac{1}{2}\left( \Phi _{i}^{s}+\Phi
_{j}^{s}\right) }\hat{\Delta}_{ij}^{\mathrm{SC}}\right\rangle \neq 0
\label{dd2}
\end{equation}%
Since the phase coherence $\left\langle e^{-i\frac{1}{2}\left( \Phi
_{i}^{s}+\Phi _{j}^{s}\right) }\right\rangle \neq 0$ is established
self-consistently once $g(i-j)$ introduces the pairing among twisted holes
as discussed above, one finally obtains $\left\langle \hat{\Delta}_{ij}^{%
\mathrm{SC}}\right\rangle \neq 0$ according to (\ref{dd2}). Note that the
phase factor $e^{-i\frac{1}{2}\left( \Phi _{i}^{s}+\Phi _{j}^{s}\right) }$
also determines the d-wave symmetry of $\left\langle \hat{\Delta}_{ij}^{%
\mathrm{SC}}\right\rangle $ as discussed previously in Ref.\cite{zhou_03}$.$

Therefore the superconducting ground state (\ref{scgs}) can be obtained as
the result of the pairing instability of twisted holes in the pseudogap
ground state (\ref{elgs}). Simultaneously the phase coherence condition (\ref%
{pc}) is achieved, and fermionic electron operators in $\hat{D}$ become
meaningful as
\begin{equation}
\hat{D}|\Psi _{G}\rangle _{SC}\simeq \sum_{ij}\tilde{g}_{ij}\sum_{\sigma
}\sigma c_{i\sigma }c_{j-\sigma }|\Psi _{G}\rangle _{SC},  \label{dd3}
\end{equation}%
with the d-wave symmetry characterized by
\begin{equation}
\tilde{g}_{ij}\simeq \frac{1}{2}\Delta _{ij}^{0}g(i-j)\left\langle e^{-i%
\frac{1}{2}\left( \Phi _{i}^{s}+\Phi _{j}^{s}\right) }\right\rangle .
\end{equation}%
Equation (\ref{dd3}) also implies the restoration of the coherent fermionic
quasiparticle excitations in the superconducting state.

\section{Summary and Discussion}

In this paper, we obtained and analyzed the ground state of a pseudogap
phase (or spontaneous vortex phase) for a doped Mott insulator, starting
from a generalized mean field treatment of the $t-J$ model in the phase
string representation. In this representation, the most singular effects of
doping, \textit{viz.}, the no double occupancy constraint, and the phase
string effect arising from the disordering of the Marshall sign, are sorted
out through a systematic procedure. All nontrivial phases in the phase
string representation are characterized by a topological gauge structure,
which turns out to be well controlled in the regime of interest (low
doping). The physical picture of the ground state that emerges, is that of
an underlying RVB vacuum (a spin liquid insulator) which makes the doped
holes be twisted into bosonic objects, thereby gaining kinetic energy. The
ground state retains time reversal and spin rotational symmetries, and has a
simple form (\ref{elgs}) in the electron second-quantization representation.

At half filling, the ground state describes an AF Mott insulator, and
provides a very accurate description of both long and short range spin
correlations of the Heisenberg model. At finite doping, the spin degrees of
freedom are described by an RVB spin liquid (\ref{rvb3}); the mean RVB pair
size (\ref{xi}) reduces continuously as doping increases. Charge degrees of
freedom emerge at finite doping, with a hidden ODLRO (\ref{odlro1}). This
leads to a finite (electron) pairing amplitude, which characterizes the
pseudogap structure of the ground state. The hidden ODLRO in the charge
sector is also found to be the underlying reason for the four nodes in the
quasiparticle excitation spectrum, but the ground state is non-Fermi-liquid
like in the absence of phase coherence. It is more appropriate to call the
pseudogap ground state as a spontaneous vortex phase, since free $S=1/2$
spins that act like vortices prevent superconducting phase coherence in this
state.

Two types of low-lying spin excitations above the pseudogap ground state
were identified. One is composed of neutral spinons and has a resonance-like
feature around ($\pi$, $\pi$). It reduces to the conventional gapless spin
wave at half-filling. The other type of spin excitation exists only in the
doped regime and describes a slowly fluctuating incommensurate AF ordering.
It provides an alternative explanation of the neutron scattering experiments
without invoking a charge stripe scenario.

Superconducting phase coherence is established by an instability which
causes RVB pairing of the free spins associated with the twisted holes. The
ensuing ground state (\ref{scgs}), is non-BCS-like in many aspects. The
properties of the superconducting ground state (\ref{scgs}), especially
those involving non-BCS-like features, are very interesting problems to
pursue further.

The class of wavefunctions discussed in this paper provides a unified
description of Mott physics, antiferromagnetism, pseudogap physics, and
d-wave superconductivity within a single framework. There are many basic
distinctions between this ground state and BCS-like wave functions. An
important distinction is the independence of the spin excitations from the
quasiparticles. The presence of the pairing amplitude and spontaneous
vortices associated with spin excitations is another unique feature and is
independent of superconducting phase coherence. Furthermore, generalized
Jastrow-like factors $\mathcal{K}$ in (\ref{k0}) that appear in the ground
state wave function, are incompatible with the (Slater) determinantal wave
functions of the BCS theory. Therefore, the pseudogap and superconducting
ground states obtained in this work represent and describe a new state of
matter.

Given that the state (\ref{elgs}) undergoes a superconducting instability
with phase coherence at low temperature, we may discuss further implications
for the global phase diagram. Clearly, the phase coherence condition (\ref%
{pc1}) fails in the spin ordered phase, when the spatial separation of the
RVB pair, $\xi \rightarrow \infty $; indeed, at half filling, the RVB
amplitude $W_{ij\text{ }}$ shows a power law decay at $\left\vert
i-j\right\vert \rightarrow \infty $, without a finite length scale, as
discussed in Sec. III A. In this AF phase, phase coherence in (\ref{pc1})
due to the tight binding of vortices and antivortices in $\Phi _{i}^{s}$ is
disrupted. In general, such a spin ordered phase will persist into finite
doping at sufficiently low hole concentration, $\delta \leq x_{c}$ \cite{kou}%
. The condition of phase coherence is not satisfied also when the RVB
pairing between spins in $|\mathrm{RVB}\rangle $ is continuously reduced to
zero, at sufficiently large doping $x_{\mathrm{RVB}}$. In this case, the
underlying rigidity for binding of vortices and antivortices in the ground
state is also destroyed and the superconductivity should be suppressed
beyond $x_{\mathrm{RVB}}$. How the ground state wavefunction systematically
evolves with the doping concentration and the possible existence of quantum
critical points at, say, $x_{c}$ and $x_{\mathrm{RVB}}$, is an important
subject for further investigation.

Finally, the results presented in this paper can be used as a starting point
to pursue several interesting questions, and we indicate a few possible
directions here: (i) The pseudogap (spontaneous vortex phase) ground state
is inherently unstable to superconductivity as the result of pairing of the
twisted holes. Nonetheless, the pseudogap ground state can still be regarded
as a true ground state, for instance, when strong magnetic fields are
applied to destroy the RVB pairing of the low-lying spins associated with
the backflow of holes. The study of this state in the presence of strong
magnetic fields is thus experimentally relevant. (ii) It will be very useful
to develop a numerical scheme based on the wavefunction (\ref{psie1}) for a
variational study. At half-filling, we used the wavefunction to compute the
ground-state energy, magnetization, and equal-time spin-spin correlations by
employing the loop gas method. As pointed out, this gave us accurate
results. However, away from half filling, the RVB amplitude $W_{ij}$ is no
longer a real function and loop products generate a fictitious flux
proportional to doping concentration [\emph{cf.} (\ref{loop})]. This leads
to a sign problem and further work is necessary to overcome this barrier.
Havilio and Auerbach studied correlations in doped antiferromagnets using a
combination of numerical methods and the Gutzwiller approximation for a
different class of RVB wave functions \cite{havilio_00}. A similar study of
the wave function proposed in this paper should yield interesting results.
(iii) Finally, we note that the form of the wave function (\ref{wavefunction}%
) follows from very general consequences of a hole moving in an RVB
background, since the disordering of the Marshall sign, \emph{i.e.}, the
phase string effect, occurs in systems even without AF long range order.
Thus the phase string representation and the bosonic RVB mean field theory
can be used to study models such as the doped quantum dimer model, where the
spin background is known to be RVB paired.

\begin{acknowledgments}
We acknowledge helpful discussions with P.W. Anderson, S.P. Kou, H.T. Nieh,
X. L. Qi, D.N. Sheng, X.G. Wen, and J. Zaanen. This work is partially
supported by the grants of the NSFC, Grant No. 104008 and SRFDP from MOE of
China.
\end{acknowledgments}

\appendix

\section{Determination of $\protect\varphi _{h}$ and $Z_{\protect\sigma }(l)$%
}

>From the mean-field state (\ref{gsph}), one finds
\begin{eqnarray}
\langle \Psi _{G}|\Psi _{G}\rangle _{\mathrm{MF}} &=&\sum_{l_{1}<l_{2}<\cdot
\cdot \cdot }\left| \varphi _{h}(\{l_{h}\})\right| ^{2}\langle \mathrm{RVB}%
|\prod_{h}Z_{\sigma _{h}^{\prime }}^{\ast }(l_{h})Z_{\sigma
_{h}}(l_{h})\left( h_{l_{h}}h_{l_{h}}^{\dagger }\right) \left(
b_{l_{h}\sigma _{h}^{\prime }}^{\dagger }b_{l_{h}\sigma _{h}}\right) |%
\mathrm{RVB}\rangle _{\mathrm{MF}}  \nonumber \\
&\approx &\sum_{l_{1}<l_{2}<\cdot \cdot \cdot }\left| \varphi
_{h}(\{l_{h}\})\right| ^{2}\prod_{h}\left| Z_{\sigma _{h}}(l_{h})\right|
^{2}\langle \mathrm{RVB}|n_{l_{h}\sigma _{h}}^{b}|\mathrm{RVB}\rangle _{%
\mathrm{MF}}\text{ .}  \label{m}
\end{eqnarray}%
In obtaining the last line, we omit RVB pairings involving two holon sites,
\textit{e.g.}, the second term on the rhs of the following expression (note
that $l_{h}$ refers to a holon site):
\begin{equation}
\langle \mathrm{RVB|}n_{l_{1}\uparrow }^{b}n_{l_{2}\downarrow }^{b}|\mathrm{%
RVB}\rangle _{\mathrm{MF}}=\langle \mathrm{RVB|}n_{l_{1}\uparrow }^{b}|%
\mathrm{RVB}\rangle _{\mathrm{MF}}\langle \mathrm{RVB}|n_{l_{2}\downarrow
}^{b}|\mathrm{RVB}\rangle _{\mathrm{MF}}+\langle \mathrm{RVB}%
|b_{l_{1}\uparrow }^{\dagger }b_{l_{2}\downarrow }^{\dagger }|\mathrm{RVB}%
\rangle _{\mathrm{MF}}\langle \mathrm{RVB}|b_{l_{1}\uparrow
}b_{l_{2}\downarrow }|\mathrm{RVB}\rangle _{\mathrm{MF}}.
\end{equation}%
The essence of this approximation is that for a dilute concentration of
holes, the hole-hole correlation induced by the background spin RVB pairing
can be neglected, since the average separation between holes sets the upper
bound for the size of the RVB pair wave function. For the sake of clarity,
we shall invoke this approximation in determining $\varphi _{h}(\{l_{h}\})$
and $Z_{\sigma _{h}}(l_{h})$ below. In principle, these effects can be
incorporated without affecting our results qualitatively.

Now, consider the hopping term (\ref{ht}), which is rewritten as
\begin{equation}
H_{t}=-t\sum_{\langle ij\rangle \sigma }\left( e^{iA_{ij}^{s}-i\phi
_{ij}^{0}-i\sigma A_{ij}^{h}}\right) \left( h_{i}^{\dagger }b_{i\sigma
}\right) \left( h_{j}b_{j\sigma }^{\dagger }\right) +h.c.~  \label{ht1}
\end{equation}%
It can be interpreted as describing the hopping of a holon-spinon composite $%
h_{i}^{\dagger }b_{i\sigma }$ under the influence of the gauge field $%
A_{ij}^{s}-\phi _{ij}^{0}-\sigma A_{ij}^{h}$ in the RVB background $|\mathrm{%
RVB}\rangle $; $b_{i\sigma }$ represents the spinon backflow accompanying
the hopping of the holon $h_{i}^{\dagger }$.

The $\uparrow $ and $\downarrow $ spinons are paired in the RVB background $|%
\mathrm{RVB}\rangle $, and it costs finite energy to break up an RVB pair at
finite doping. Correspondingly, $\left\langle A_{ij}^{s}\right\rangle =0$
and $\left\langle \left( A_{ij}^{s}\right) ^{2}\right\rangle $ remains small
in (\ref{ht1}) because $A_{ij}^{s}$ depicts fictitious $\pm \pi $ fluxoids
bound to $\sigma =\pm 1$ spins. We point out that this condition could fail
both, at very low doping, ($\delta <x_{c}$ \cite{kou}), where a spin ordered
phase exists, and at large doping ($\delta >x_{\mathrm{RVB}}$), where the
RVB pairing disappears and $A_{ij}^{s}$ fluctuates strongly. In both limits,
we expect different phases.

Neglecting $A_{ij}^s$ and the hard core repulsion among holons, one may
simply choose
\begin{equation}
\varphi _h(l_1,l_{2,}...,l_{N_h})=\text{\textrm{cons}}\mathrm{t.}
\label{aph}
\end{equation}
as if the holons are in a Bose condensed state. Consistent with this picture
of holon condensation, $A_{ij}^h$ can indeed be treated as describing a
constant flux $\pi \delta $ per plaquette. We recall that such an
approximation [\textit{cf.}(\ref{fluxh})] was made in determining the RVB
amplitudes, $W_{ij}$.

We can now determine $Z_{\alpha _h}(l_h)$ by minimizing $\langle
\Psi_G|H_t|\Psi _G\rangle _{\mathrm{MF}}/\langle \Psi _G |\Psi _G\rangle _{%
\mathrm{MF}}$. Upon invoking the approximations detailed above, we get
\begin{equation}
\frac{\langle \Psi _G|H_t|\Psi _G\rangle _{\mathrm{MF}}}{\langle \Psi
_G|\Psi _G\rangle _{\mathrm{MF}}}\simeq -\ \tilde{t}\sum_{h=1}^{N_h}%
\sum_{<ij>}e^{-i\phi _{ij}^0-i\sigma _hA_{ij}^h}Z_{\sigma
_h}^{*}(i)Z_{\sigma _h}(j)~,
\end{equation}
where
\begin{equation}
\tilde{t}=\left( \frac{\bar{n}^b}2+\frac{\left| \Delta ^s\right| ^2}{2\bar{n}%
^b}\right) t,\text{ \qquad }\bar{n}^b=1-\delta ~.  \label{ath}
\end{equation}
By optimizing the hopping energy under the condition $\sum_i\left| Z_{\sigma
_h}(i)\right| ^2=1,$ we get the hopping energy
\begin{equation}
\langle \Psi _G|H_t|\Psi _G\rangle _{\mathrm{MF}}/\langle \Psi _G|\Psi
_G\rangle _{\mathrm{MF}}=-\ 4t_h\delta N~,
\end{equation}
with $t_h>0$ being the maximal eigenvalue of the eigen equation
\begin{equation}
(-t_h)Z_{\sigma _h}(i)=-\frac{\tilde{t}}4\sum_{j=nn(i)}e^{-i\phi
_{ij}^0-i\sigma _hA_{ij}^h}Z_{\sigma _h}(j)~,  \label{azeq}
\end{equation}
and $Z_{\sigma _h}$, its eigen wave function.

>From (\ref{ath}) we see that $\tilde{t}$ has two contributions: the first
term, proportional to $(1-\delta )/2,$ is the probability that the hole hops
between nn sites, in a background of uncorrelated spins; the second term,
proportional to $|\Delta ^{s}|^{2},$ is the enhancement from RVB assisted
hopping - the initial state (before hole hopping) and the final state have a
non vanishing overlap owing to the RVB pairing in the spin background. Note
that $\tilde{t}\simeq t$ at small $\delta $ as $\bar{n}^{b}\sim 1,$ $\Delta
^{s}\sim 1$. The effective hopping integral $t_{h}$ can be obtained from (%
\ref{azeq}), for a fixed hole concentration and under the approximation (\ref%
{fluxh}). The numerical results are shown in Fig. 4.

Finally, we note that the eigen values of (\ref{azeq}) would form a
Hofstadter spectrum, if the gauge field $A_{ij}^h$ were treated at a mean
field level, as a uniform flux (\ref{fluxh}). However, the density of the
backflow spinons is tied to that of the holons and so the fluctuations in $%
A_{ij}^h$ due to fluctuations in the holon density are self consistently
related to the density fluctuations of the spinon backflow. Thus, (\ref{azeq}%
) resembles an anyon (semion) system in the bosonic representation with a
spin index \cite{wen_90}. The mean field approximation is usually no longer
valid in determining its excitation spectrum. This is a nontrivial issue.
Nevertheless, since we shall be only interested in its ground state, which
is a condensate with a uniform density distribution, all we need to know is
that there is a renormalized constant $t_h$ for the effective holon hopping
term, estimated as the maximal eigen-value at the mean-field level in (\ref%
{azeq}). The wave packet $Z_\sigma (i)$ is determined as a linear
combination of the wave packets of a cyclotron radius $a_c=a/\sqrt{\pi
\delta }$ in accord with (\ref{fluxh}) and is made maximally uniform in
space, in order to accommodate the condensation of the holes.

\section{Effective theory}

In the bosonic RVB mean field theory, the holon many-body wave function $%
\varphi _{h}$ has been assumed to satisfy the ideal Bose condensation
condition (\ref{hcond}) at $T=0$. As we noted in the paper, two effects are
omitted in (\ref{hcond}), \textit{viz.}, the hard core correlation among
holons and the fluctuations of the link field $A_{ij}^{s}$. The latter
effect should become important when excited spinons are present.
Furthermore, $\varphi _{h}$ will also become nontrivial in the presence of
an external electromagnetic field.

A general form of the wave function $\varphi _h$ in (\ref{gsph}) under the
influence of $A_{ij}^s$, and in the presence of an external electromagnetic
field $A_{ij}^e$, can be determined by the following effective hopping term
\begin{equation}
H_h=-\ t_h\sum_{\langle ij\rangle }e^{iA_{ij}^s+iA_{ij}^e}h_i^{\dagger
}h_j+h.c.  \label{hh}
\end{equation}
If the hard core condition of the holon field is neglected, and if $A_{ij}^e
\equiv0$, (\ref{hh}) leads to the Bose condensation solution (\ref{hcond})
in the ground state. Note that $H_h$ is a gauge model consistent with the
gauge invariance (\ref{u(1)1}) in the original Hamiltonian. $H_h$ also
respects the spin rotational symmetry as one can easily check that $\left[
H_h,\mathbf{S}\right] =0$, where $\mathbf{S}$ is the total spin operator.
Furthermore, in the absence of $A_{ij}^e$, the time reversal symmetry of $%
H_h $ can also be shown, by noting that $A_{ij}^s\rightarrow -A_{ij}^s$
under the flip of the spins.

The effective hopping integral $t_h$ appears in (\ref{hh}) as a renormalized
$t$, which is decided by the spinon backflow according to (\ref{zeq}). Here
the spinons do not directly see the gauge field $A_{ij}^s$ and the external
electromagnetic field $A_{ij}^e$, because they satisfy a different gauge
transformation (\ref{u(1)2}) in (\ref{zeq}). Thus, the spinons are truly
charge neutral and only carry $S=1/2$ in the phase string formalism, in
contrast to, say, the slave boson gauge theories where both holons and
spinons are coupled to the same external electromagnetic field through the
Ioffe-Larkin rule.

An effective Hamiltonian $H_{\mathrm{string}}$ based on $H_{h}$ and $H_{s}$
can be written down as
\begin{equation}
H_{\mathrm{string}}=H_{h}+H_{s}~.  \label{hstring}
\end{equation}%
Such an effective Hamiltonian has been derived earlier \cite{weng_99}, and
is the basic low energy effective model for the doping and temperature
regimes underpinned by the bosonic RVB order parameter $\Delta ^{s}\neq 0$
in the phase diagram of Fig. 2.

We remark that $H_{\mathrm{string}}$ in (\ref{hstring}) is obtained under
the assumption that the spinon backflow accompanying the hopping of the
holons only provides a renormalized hopping integral $t_h$. Here $t_h$ is
determined at the mean field level, as the minimal eigen value of (\ref{zeq}%
). However, in principle, the internal excitations of the spinon backflow
can effectively reduce $t_h$ according to (\ref{zeq}) and thus increase the
kinetic energy of $H_{\mathrm{string}}$. This should be taken into account
when the high energy part of the charge degrees of freedom are studied.
Finally, we point out that the holon and its spinon backflow is actually
bound together, as previously shown \cite{zhou_03} by including the residual
interactions of the $t-J$ model beyond the effective Hamiltonian. Such a
twisted hole as a bound holon-spinon pair emerges naturally in the present
wavefunction approach given in the main text.

\section{Time reversal and spin rotational symmetries of the pseudogap
ground state (\protect\ref{elgs})}

\subsection{Time reversal symmetry}

The time reversal relation for an $S=1/2$ single-particle wavefunction is

\begin{equation}
\psi _{\sigma }^{T}=\sigma \psi _{-\sigma }^{\ast }.
\end{equation}%
Correspondingly, in second quantized language, the time reversal of the
electron operators read

\begin{eqnarray}
Tc_{i\sigma }^{\dag }T^{-1} &=&\sigma c_{i-\sigma }^{\dag }, \\
Tc_{i\sigma }T^{-1} &=&\sigma c_{i-\sigma }.
\end{eqnarray}%
It is easy to check that
\begin{equation}
T\mathbf{S}_{i}T^{-1}=-\mathbf{S}_{i}.
\end{equation}%
Then the time reversal of the RVB state $|\mathrm{RVB}\rangle $ is given by
\begin{eqnarray}
T|\mathrm{RVB}\rangle &=&\mathrm{const.}\sum_{\{\sigma _{s}\}}\Phi _{\mathrm{%
RVB}}^{\ast }(\sigma _{1},\sigma _{2},\cdot \cdot \cdot \sigma _{N})\left[
\sigma _{1}\sigma _{2}\cdot \cdot \cdot \sigma _{N}\right] c_{1-\sigma
_{1}}^{\dagger }c_{2-\sigma _{2}}^{\dagger }\cdot \cdot \cdot c_{N-\sigma
_{N}}^{\dagger }|0\rangle  \nonumber \\
&=&\mathrm{const.}\sum_{\{\sigma _{s}\}}\Phi _{\mathrm{RVB}}^{\ast }(-\sigma
_{1},-\sigma _{2},\cdot \cdot \cdot ,-\sigma _{N})c_{1\sigma _{1}}^{\dagger
}c_{2\sigma _{2}}^{\dagger }\cdot \cdot \cdot c_{N\sigma _{N}}^{\dagger
}|0\rangle  \nonumber \\
&=&|\mathrm{RVB}\rangle
\end{eqnarray}%
by noting that $\sigma _{1}\sigma _{2}\cdot \cdot \cdot \sigma
_{N}=(-1)^{N/2}$ in a bipartite lattice$,$ and
\begin{eqnarray}
\Phi _{\mathrm{RVB}}^{\ast }(-\sigma _{1},-\sigma _{2},\cdot \cdot \cdot
,-\sigma _{N}) &=&\sum_{\mathrm{pair}}\prod_{(ij)}(-1)^{i}(-1)W_{ij}^{\ast }
\nonumber \\
&=&\sum_{\mathrm{pair}}\prod_{(ij)}(-1)^{i}(-1)W_{ji}  \nonumber \\
&=&\sum_{\mathrm{pair}}\prod_{(ij)}(-1)^{j}W_{ji}  \nonumber \\
&=&\Phi _{\mathrm{RVB}}(\sigma _{1},\sigma _{2},\cdot \cdot \cdot ,\sigma
_{N}).
\end{eqnarray}%
Furthermore, by using

\begin{eqnarray*}
Te^{-i\hat{\Theta}_{i\sigma }}T^{-1} &=&e^{i\frac{1}{2}\left[ -\Phi
_{i}^{s}-\Phi _{i}^{0}-\sigma \Phi _{i}^{h}\right] }~(\sigma )^{\hat{N}%
_{h}}(-\sigma )^{i} \\
&=&e^{-i\hat{\Theta}_{i-\sigma }}e^{-i\Phi _{i}^{0}}(-1)^{\hat{N}%
_{h}}(-1)^{i}
\end{eqnarray*}%
according to Eq.(\ref{phase}) and $\tilde{c}_{i\sigma }=e^{-i\hat{\Theta}%
_{i\sigma }}c_{i\sigma }$, one finds that
\begin{equation}
T\tilde{c}_{i\sigma }T^{-1}=\sigma \tilde{c}_{i-\sigma }e^{-i\Phi
_{i}^{0}}(-1)^{\hat{N}_{h}}(-1)^{i}.
\end{equation}%
Finally, the time reversal of the ground state $|\Psi _{G}\rangle $ is given
by

\begin{eqnarray*}
T|\Psi _{G}\rangle &=&T\left( \sum_{l}Z_{\uparrow }(l)\tilde{c}_{l\uparrow
}\right) ^{N_{h}^{\uparrow }}\left( \sum_{l^{\prime }}Z_{\downarrow
}(l^{\prime })\tilde{c}_{l^{\prime }\downarrow }\right) ^{N_{h}^{\downarrow
}}T^{-1}T|\mathrm{RVB}\rangle \\
&=&\left( \sum_{l}Z_{\uparrow }^{\ast }(l)\tilde{c}_{l\downarrow }e^{-i\Phi
_{l}^{0}}(-1)^{\hat{N}_{h}}(-1)^{l}\right) ^{N_{h}^{\uparrow }}\left(
-\sum_{l^{\prime }}Z_{\downarrow }^{\ast }(l^{\prime })\tilde{c}_{l^{\prime
}\uparrow }e^{-i\Phi _{l^{\prime }}^{0}}(-1)^{\hat{N}_{h}}(-1)^{l^{\prime
}}\right) ^{N_{h}^{\downarrow }}|\mathrm{RVB}\rangle \\
&=&(-1)^{N_{h}^{\downarrow }}(-1)^{N_{h}(N_{h}-1)/2}\left(
\sum_{l}Z_{\downarrow }(l)\tilde{c}_{l\downarrow }\right) ^{N_{h}^{\uparrow
}}\left( \sum_{l^{\prime }}Z_{\uparrow }(l^{\prime })\tilde{c}_{l^{\prime
}\uparrow }\right) ^{N_{h}^{\downarrow }}|\mathrm{RVB}\rangle \\
&=&|\Psi _{G}\rangle
\end{eqnarray*}%
at $N_{h}^{\uparrow }=N_{h}^{\downarrow }=N_{h}/2$ (or $S^{z}=0$). In
obtaining the third line, the condition (\ref{zd}) is used.

Thus, we have proved the time reversal symmetry of the pseudogap ground
state (\ref{elgs}), for $S^z=0$.

\subsection{Spin rotational symmetry}

The condition for a state with $SU(2)$ spin rotational symmetry is

\begin{equation}
\mathbf{S}^2\left| \Psi \right\rangle =0
\end{equation}
or equivalently,

\begin{eqnarray}
S^z|\Psi \rangle &=&0, \  \\
S^{\pm }|\Psi \rangle &=&0.
\end{eqnarray}
Here $\mathbf{S}$ is the total spin and $S^{\pm }=S^x\pm iS^y$, where $S^x$,
$S^y$ and $S^z$ are the three components of $\mathbf{S}$. In the following
we shall consider $|\Psi _G\rangle $ with $S^z=0,$ and show that $S^{\pm
}|\Psi _G\rangle =0.$

By noting that, for $i\neq l$,
\begin{eqnarray}
S_i^{+}e^{-i\frac 12\Phi _l^s} &=&e^{-i\frac 12\Phi _l^s}S_i^{+}e^{i\theta
_l(i)} \\
S_i^{+}e^{i\frac \sigma 2\Phi _l^h} &=&e^{i\frac \sigma 2\Phi _l^h}S_i^{+}
\end{eqnarray}
we have
\begin{eqnarray}
S_i^{+}\tilde{c}_{l\sigma } &=&S_i^{+}e^{-i\hat{\Theta}_{l\sigma
}}c_{l\sigma }  \nonumber \\
&=&e^{i\theta _l(i)}\tilde{c}_{l\sigma }S_i^{+}(1-\delta _{il})
\end{eqnarray}
and
\begin{equation}
S^{+}\tilde{c}_{l\sigma }=\tilde{c}_{l\sigma }\sum_{i\neq
l}S_i^{+}e^{i\theta _l(i)}.
\end{equation}
Then according to (\ref{elgs0}) and (\ref{hcond}),
\begin{eqnarray}
S^{+}|\Psi _G\rangle &=&S^{+}\sum_{\{l_h\}}\prod_hZ_{\sigma _h}(l_h)\tilde{c}%
_{l_h\sigma _h}|\mathrm{RVB}\rangle  \nonumber \\
&=&\sum_{\{l_h\}}\prod_hZ_{\sigma _h}(l_h)\tilde{c}_{l_h\sigma
_h}\left(\sum_{i\neq l_h}S_i^{+}e^{i\sum_{l_h\neq i}\theta _{l_h}(i)}|%
\mathrm{RVB}\rangle \right) .  \label{rvb2}
\end{eqnarray}

In terms of (\ref{rvb3}) and (\ref{phirvb-1}), one has
\begin{eqnarray}
|\mathrm{RVB}\rangle &=&\sum_{\{\sigma _s\}}\Phi _{\mathrm{RVB}}(\sigma
_1,\sigma _2,\cdot \cdot \cdot \sigma _N)c_{1\sigma _1}^{\dagger }c_{2\sigma
_2}^{\dagger }\cdot \cdot \cdot c_{N\sigma _N}^{\dagger }|0\rangle  \nonumber
\\
&\equiv &\mathrm{const.}\sum_{\{\sigma _s\}}\sum_{\mathrm{pair}%
}\prod_{(ij)}(-1)^iW_{ij}\left| \cdot \cdot \cdot ,i\uparrow ,j\downarrow
,\cdot \cdot \cdot \right\rangle ,  \label{rvb4}
\end{eqnarray}
where each spin configuration $\{\sigma _1,\sigma _2,\cdot \cdot \cdot
\sigma _N\}$ is partitioned in pairs denoted by $(ij)$, with $i$ and $j$
belonging to different sublattices connected by $(-1)^iW_{ij}$ in (\ref{rvb4}%
). For each pair of $(ij)$ and $(ji)$, the rest of $|\mathrm{RVB}\rangle$
remains the same (the nature of RVB pairing) and thus needs not to be
considered. Then we find
\begin{eqnarray}
I_{ij} &\equiv &\left( S_i^{+}e^{i\sum_{l_h\neq i}\theta
_{l_h}(i)}+S_j^{+}e^{i\sum_{l_h\neq j}\theta _{l_h}(j)}\right) \left[
(-1)^iW_{ij}\left| \cdot \cdot \cdot ,i\uparrow ,j\downarrow ,\cdot \cdot
\cdot \right\rangle +(-1)^jW_{ji}\left| \cdot \cdot \cdot ,i\downarrow
,j\uparrow ,\cdot \cdot \cdot \right\rangle \right]  \nonumber \\
&=&\left( (-1)^je^{i\sum_{l_h\neq i}\theta
_{l_h}(i)}W_{ji}+(-1)^ie^{i\sum_{l_h\neq j}\theta _{l_h}(j)}W_{ij}\right)
\left| \cdot \cdot \cdot ,i\uparrow ,j\uparrow ,\cdot \cdot \cdot
\right\rangle  \nonumber \\
&=&e^{\frac i2\left( \sum_{l_h\neq i}\theta _{l_h}(i)+\sum_{l_h\neq j}\theta
_{l_h}(j)\right) }(-1)^j\left(
W_{ji}e^{iA_{ij}^h}-W_{ij}e^{-iA_{ij}^h}\right) \left| \cdot \cdot \cdot
,i\uparrow ,j\uparrow ,\cdot \cdot \cdot \right\rangle  \label{I}
\end{eqnarray}
where in obtaining the last line, $\theta _{l_h}(i)-\theta _{l_h}(j)=\theta
_i(l_h)-\theta _j(l_h)$ has been used.

In the ground state, the holes are uniformly distributed in space due to the
ODLRO, which ensures (\ref{fluxh}). We shall show (see below) that under the
condition (\ref{fluxh}) one has the following relation,
\begin{equation}
W_{ij}=|W_{ij}|e^{iA_{ij}^{h}}  \label{wij2}
\end{equation}%
up to a pure gauge transformation. Since $|W_{ij}|=|W_{ji}|,$ then one has
\begin{equation}
I_{ij}=0  \label{Iij}
\end{equation}%
for any given set of RVB pair $(ij)$ in (\ref{rvb4}).

Note that one should be careful about the case that a pair of sites $(ij)$
may be occupied by holes. Obviously there is no contribution in (\ref{rvb2})
if two sites are both occupied by holes, or by a hole and an up spin. If
only one of them is occupied by a hole and the other is by a down spin, then
two configurations, with the hole at site $i$/$j$ and a down spin at site $j$%
/$i$, will have the same amplitude but opposite signs similar to the case in
(\ref{I}). Since the twisted holes are Bose condensed, smearing out the
charge distribution one still finds a cancellation with vanishing
contribution to (\ref{rvb2}). Therefore, generally one has
\begin{equation}
S^{+}|\Psi _{G}\rangle =0.
\end{equation}

\emph{Proof of (\ref{wij2}) under the condensation condition (\ref{fluxh}).}
According to the definition (\ref{wij}), $W_{ij}$ is determined by (\ref{ew}%
), which is rewritten as
\begin{equation}
\sum_{j=NN(i)}e^{-iA_{ji}^{h}}w_{m}(j)=\xi _{m}^{\prime }w_{m}(i)
\label{ew1}
\end{equation}%
with $w_{m}\equiv w_{m\uparrow }(i)$ and $\xi _{m}^{\prime }\equiv -\xi
_{m}/J_{s},$ with a gauge choice
\begin{eqnarray}
A_{i,i+\hat{x}}^{h} &=&0,\  \\
A_{i,i+\hat{y}}^{h} &=&\left( \pi \delta \right) i_{x}  \label{fluxh1}
\end{eqnarray}%
for (\ref{fluxh}) under the ``holon'' condensation condition. Since the
system is translational invariant in the $\hat{y}$-direction, we may express
$w_{m}(i)$ as $%
w_{m}(i)=e^{im_{y}i_{y}}g_{m}(i_{x})=e^{im_{y}i_{y}}g_{m_{x},m_{y}}(i_{x}),$
with the eigenfunction (\ref{ew1}) becomes
\begin{equation}
g_{m}(i_{x}-1)+g_{m}(i_{x}+1)+2\cos (\pi \delta i_{x}-m_{y})g_{m}(i_{x})=\xi
_{m}^{\prime }g_{m}(i_{x}),  \label{ew3}
\end{equation}%
where $m_{y}=2\pi n/L$, $n=0,1,...,L-1$. If the doping concentration $\delta
$ satisfies $\pi \delta L=2k\pi $ ($k\in \mathcal{Z}$), the periodical
boundary condition can be chosen (the Hofstadter case). The more general
case can always be infinitesimally approached from the Hofstadter case.

In terms of (\ref{ew3}), we have the following identities
\begin{equation}
g_{m_x,m_y}(i_x+1)=g_{m_x,m_y-\pi \delta }(i_x)e^{i\chi _m}  \label{ew4}
\end{equation}
and
\begin{equation}
\xi _{m_x,m_y}^{\prime }=\xi _{m_x,m_y-\pi \delta }^{\prime },  \label{ew5}
\end{equation}
where $\chi _m$ depends on the choice of the phase of $g_m$. Using the above
identities, we obtain
\begin{equation}
g_{m_x,m_y}(i_x+\eta _x)=g_{m_x,m_y-\pi \delta \eta
_x}(i_x)e^{i\sum_{n=1}^{\eta _x}\chi _{m+(n-1)\hat{x}}}.
\end{equation}
Substitute the above solution into $W_{ij}$,
\begin{eqnarray*}
W_{ij} &=&-\sum_m\frac{v_m}{u_m}w_m(i)w_m^{*}(j) \\
&=&-\sum_m\frac{v_m}{u_m}%
e^{im_y(i_y-j_y)}g_{m_x,m_y}(i_x)g_{m_x,m_y}^{*}(j_x).
\end{eqnarray*}
Then
\begin{eqnarray*}
W_{i+\eta ,j+\eta } &=&-\sum_m\frac{v_m}{u_m}%
e^{im_y(i_y-j_y)}g_{m_x,m_y}(i_x+\eta _x)g_{m_x,m_y}^{*}(j_x+\eta _x) \\
&=&-\sum_m\frac{v_m}{u_m}e^{im_y(i_y-j_y)}g_{m_x,m_y-\pi \delta \eta
_x}(i_x)g_{m_x,m_y-\pi \delta \eta _x}^{*}(j_x) \\
&=&-\sum_m\frac{v_m}{u_m}e^{i(m_y+\pi \delta \eta
_x)(i_y-j_y)}g_{m_x,m_y}(i_x)g_{m_x,m_y}^{*}(j_x) \\
&=&e^{i\pi \delta \eta _x(i_y-j_y)}W_{ij}.
\end{eqnarray*}
and it is straightforward to show
\begin{equation}
\prod_{\text{loop}}W_{ij}=\prod_{\text{loop}}|W_{ij}|\cdot e^{i\sum_{\text{%
loop}}A_{ij}^h},  \label{loop}
\end{equation}
on a loop $(i\rightarrow j\rightarrow j+\eta \rightarrow i+\eta \rightarrow
i)$. One can always choose a proper gauge such that (\ref{wij2}) holds on
any link $(i,j)$.


\begin{thebibliography}{99}
\bibitem{remark1} Note that the naive form of $|\mathrm{GS}\rangle =|\Psi
\rangle _{h}\otimes |\mathrm{RVB}\rangle _{\mathrm{MF}}$ does not optimize
the kinetic energy of the hole. This can be seen most directly by
calculating $\langle \mathrm{GS}|H_{t}|\mathrm{GS}\rangle $, where $H_{t}$
is the hole hopping term in the $t-J$ Hamiltonian, (\ref{ht}). Since $%
\langle b_{i\sigma }^{\dagger }b_{i\sigma }\rangle _{nn}\equiv 0$, the
kinetic energy of the hole vanishes. In the earlier approach \cite{weng_99},
to avoid this difficulty, a hidden symmetry in $|\mathrm{RVB}\rangle _{%
\mathrm{MF}}$ has to be used to get a finite and optimized $\langle \mathrm{%
GS}|H_{t}|\mathrm{GS}\rangle ,$ which makes $|\mathrm{RVB}\rangle _{\mathrm{%
MF}}$ itself also explicitly depend on the hole configuration.
\bibitem{remark3} With the increase of the hole concentration, the
\textquotedblleft hard-core\textquotedblright\ condition among bosonic $%
\tilde{c}_{l\sigma }$'s will gradually become important and eventaully the
above \textquotedblleft ideal\textquotedblright\ Bose condensation
approximation has to be modified.
\bibitem{remark4} This kind of spin excitation involving the twisted holes
has some similarity to the usual d-wave BCS theory as in both cases the spin
excitation is made of nodal quasiparticles. However, there will be an
important distinction since quasiparticles in the present case are no longer
elementary in the conventional sense as pointed out in the text.
\end{thebibliography}
\end{document}